\documentclass[pre,showpacs,reprint,amsmath,amssymb,superscriptaddress]{revtex4-2}

\usepackage{amsmath}
\usepackage{wasysym}
\usepackage{graphicx}
\usepackage[OMLmathrm, rmdefault=mdput]{isomath}
\usepackage{upgreek}
\usepackage{xcolor}
\usepackage{bm}
\usepackage{mathtools}

\usepackage[normalem]{ulem}

\newcommand*{\defeq}{\mathrel{\vcenter{\baselineskip0.5ex \lineskiplimit0pt
                     \hbox{\small.}\hbox{\small.}}}%
                     =}

\newcommand{\dd}{d}
\newcommand{\Tr}{\mathop{\textrm{Tr}}}
\newcommand{\p}{n}
\newcommand{\x}{X}
\renewcommand{\u}{{}}
\newcommand{\uu}{{\mathfrak{u}\hspace{.1ex}}}
\newcommand{\n}{m}
\renewcommand{\o}{i}
\newcommand{\ov}{j}
\newcommand{\obar}{{\bar{\if i\o \imath \else\if j\o \jmath \else \o \fi\fi}}}
\newcommand{\ovbar}{{\bar{\if i\ov \imath \else\if j\ov \jmath \else \ov \fi\fi}}}
\newcommand{\s}{{\text{k}}}
\newcommand{\sv}{{\text{l}}}

\newcommand{\so}{{\text{\o}}}
\newcommand{\sov}{{\text{\ov}}}
\newcommand{\sobar}{{\bar{\mathrm{\if i\o \text{\i} \else\if j\o \text{\j} \else \o \fi\fi}}}}
\newcommand{\sovbar}{{\bar{\mathrm{\if i\ov \text{\i} \else\if j\ov \text{\j} \else \ov \fi\fi}}}}
\newcommand{\mysum}[1]{\sum_{#1}}
\newcommand{\myprod}[1]{\prod_{#1}}

\newcommand{{\nul}}{}
\newcommand{\ovec}{\boldsymbol{\omega}}
\newcommand{\svec}{\mathrm{\omega}}

\begin{document}

\title{Ordering and association of patchy particles in quasi-one-dimensional channel}

\author{Péter Gurin}
\email{peter.gurin@mk.uni-pannon.hu}
\affiliation{ Physics Department, Centre for Natural Sciences, University of Pannonia, PO Box 158, Veszprém, H-8201 Hungary}
\author{Szabolcs Varga}
\email{szabolcs.varga@mk.uni-pannon.hu}
\affiliation{ Physics Department, Centre for Natural Sciences, University of Pannonia, PO Box 158, Veszprém, H-8201 Hungary}

\begin{abstract}
We show that the formalism of Wertheim's first order thermodynamic perturbation theory can be generalised for the fluid of
rotating sticky particles with anisotropic hard core confined to a quasi-one-dimensional channel. Using the transfer matrix method, we prove that the theory is exact if the hard body interaction is additive, only the first neighbors interact and the particles can stick together only along the channel. We show that the most convenient treatment of association in narrow channels is to work in NPT ensemble, where all structural and thermodynamic quantities can be expressed as a function of pressure and fraction of sites unbonded.
\end{abstract}

\maketitle


\section{Introduction}

Self-assembly of patchy particles has been a hot topic for a while due
to many practical applications and fundamental issues of condensed
matter~\cite{%
  Glotzer-Solomon_NatureMater_2007,%
  Zhang-Regulacio-Han_ChemSocRev_2014
}. One key issue is how to synthesise anisotropic building blocks so
that they self-organise into a target structure~\cite{%
  Coluzza-vanOostrum-Capone-Reimhult-Dellago_SoftMatter_2013
}. It is also unclear how the confinement affects the
self-organisation of the building blocks~\cite{%
  Leferink.opReinink-van.denPol-Petukhov-Vroege-Lekkerkerker_EurPhysJSpecTop_2013,%
  Royall-Charbonneau-Dijkstra-Russo-Smallenburg-Speck-Valeriani_RevModPhys_2024%
}. Theory and simulation can be very helpful in answering these
questions, because they can provide useful results starting from
particle-particle and particle-confinement interactions. However these
interactions are anisotropic due to orientation dependent features of
the system such as the molecular shape, the directional bonding of the
patches and the geometry of confinement~\cite{%
  Jack-Millett_AIPAdvances_2021,%
  Kalyuzhnyi-Patsahan-Holovko-Cummings_Nanoscale_2024%
}. To overcome the time-consuming calculations, it is advisable to
consider toy models that can be solved exactly or the effects of
different properties can be well decoupled from each other
\cite{Vo-Glotzer_PNAS_2022}.

It is well-known that one-dimensional (1D) systems provide the best
playground to get insight into the structural and thermodynamic
properties of strongly confined two-, and three-dimensional fluids,
because they mimic more or less the properties of realistic systems
and can often be studied exactly using statistical mechanics~\cite{%
  Herzfeld-GoeppertMayer_JChemPhys_1934,%
  Brader-Evans_PhysicaA_2002,%
  Fantoni-Santos_JStatPhys_2017,%
  Montero-Santos_JStatPhys_2019,%
  Meddour-Bouzar-Messina_JPhysCondensMatter_2025%
}. Particles in 1D confinement can form vapour, liquid and even solid
phases if the pair interaction is very long ranged
\cite{Kac_PhysFluids_1959} and the applied external field is periodic
\cite{Carraro_PhysRevE_2003}. The effect of association can be studied
exactly in some patchy systems, such as the 1D fluid of two-faced
Janus particles~\cite{%
  Fantoni-Maestre-Santos_JStatMech_2021%
}. It is also possible to study the orientational ordering~\cite{%
  Lebowitz-Percus-Talbot_JStatPhys_1987,%
  Kantor-Kardar_EPL_2009
}, nucleation~\cite{Joswiak-Doherty-Peters_JChemPhys_2016}, glass
formation~\cite{%
  Bowles_PhysicaA_2000,%
  Godfrey-Moore_PhysRevE_2015%
} and the jamming~\cite{%
  Ashwin-Bowles_PRL_2009,%
  Zarif-Spiteri-Bowles_PhysRevE_2021%
} in some quasi-one-dimensional (q1D) systems. It turned out that the
most powerful methods to study q1D systems is the Laplace
transformation method~\cite{Salsburg-Zwanzig-Kirkwood_JCP_1953}, the
transfer matrix method (TMM)~\cite{%
  Casey-Runnels_JChemPhys_1969,%
  Kofke-Post_JCP_1993%
}, the neighbour distribution method
\cite{Santos_LectureNotesInPhys_2016} and the classical density
functional theory (DFT) \cite{Percus_MolPhys_2002}. Among these
methods, DFT is the most widely used method for the investigation of
thermodynamic and structural properties of inhomogeneous liquids,
liquid crystals and solids, as it can be extended to higher
dimensional systems~\cite{%
  Mulero_LectureNotesInPhys_2008,%
  Luis-Enrique-Yuri_JPhysCondMat_2014%
}. The development of it is highly due to the 1D model systems for
which the bulk and structural properties can be determined exactly if
pair interaction is very short or very long ranged~\cite{%
  Herzfeld-GoeppertMayer_JChemPhys_1934
}. The famous system is the 1D fluid of hard rods, where the particles
are not allowed to overlap with each other and to rotate out of the
confining line. 
This was the first system
for which exact DFT is developed in the presence of arbitrary external
field thanks to J. K. Percus \cite{Percus_JStatPhys_1976}. Later, he
derived an exact density functional for sticky hard rod, where both
ends of the rod particle are decorated by attractive patches~\cite{%
  Percus_JStatPhys_1982}. Along this line, Kierlik and Rosinberg
extended the list of exact density functionals with the dimerizing and
chain forming sticky hard rods, where the rods can flip between two
orientations~\cite{Kierlik-Rosinberg_JStatPhys_1982}. Later, the above
DFTs were generalised for sticky hard rod mixtures~\cite{%
  Vanderlick-Davis-Percus_JChemPhys_1989,%
  Brannock-Percus_JChemPhys_1996,%
  Percus_JStatPhys_1997,%
  Tutschka-Cuesta_JStatPhys_2003%
}. Noteworthy those exact DFTs derived for some rod systems moving on
a 1D lattice~\cite{%
  Sahnoun-Djebbar-Benmessabih-Bakhti_JPhysA-MathTheor_2024%
}, which mimic the form of the well-known fundamental measure theory
(FMT) used mainly in higher dimensions~\cite{Rosenfeld_PRL_1989}.
Apart from the above list, no exact density functionals are available
for several 1D fluids such as the well-known 1D Lennard-Jones fluid,
for which approximate DFTs were found using machine learning
technique~\cite{Lin-Oettel_SciPostPhys_2019
}. In the framework of FMT, several DFT were devised to describe the
structural properties of bulk and confined spherical and non-spherical
patchy particles~\cite{Wu-Li_AnnuRevPhysChem_2007}. However, even the
1D binary mixture of non-additive hard rods cannot be described
exactly using the FMT~\cite{%
  Schmidt_PRE_2007
}. In addition, the out-of-line orientational freedom gives extra
difficulty in the application of FMT~\cite{%
  ElMoumane-teVrugt-Loewen-Wittmann_JChemPhys_2024%
}, which means that only approximate density functionals are available
even for studying freely rotating patchy hard spheres moving on a line
\cite{Marshall_JChemPhys_2015,Marshall_PhysRevE_2016}.

One possible solution to overcome the weakness of the DFT is to use the TMM, which is an exact method for the calculation of the isobaric partition function if only few (first, second, etc.) neighbour interactions are present~\cite{Percus-Zhang_MolPhys_1990,Gurin-Varga_JCP_2015}. Therefore, it can be used for several single-file fluids such as the colloidal particles in cylindrical nanopores~\cite{
  Lin-Valley-Meron-Cui-Ho-Rice_JPhysChemB_2009
} and fullerenes encapsulated into carbon nanotube~\cite{
  Gorantla_et.al_Nanoscale_2010}. Along this line, there are two main classes of TMM applications: 1) particles interacting with isotropic pair potentials in a narrow channel~\cite{Kofke-Post_JCP_1993} and 2) particles interacting with anisotropic pair potentials confined to a straight line~\cite{Lebowitz-Percus-Talbot_JStatPhys_1987}. These two classes are of the same level of complexity, as only the out-of-line positional freedoms need to be replaced by orientational ones when moving from the first to the second class. However, the only common in these two classes is that they are both q1D systems. The differences can be summarised briefly as follows. For example, the q1d system of hard spheres in narrow channel undergoes a fluid-solid like structural change with emerging jamming, glass formation and caging phenomena~\cite{Ashwin-Yamchi-Bowles_PhysRevLett_2013,Robinson-Godfrey-Moore_PhysRevE_2016
}, while the rod-like particles on a straight line exhibit peculiar close packing behaviour with diverging orientational and positional correlation lengths~\cite{Montero-Santos-Gurin-Varga_JCP_2023,Gurin-Mizani-Varga_PhysRevE_2024}. Only recently it has become clear that orientational disorder-order structural change is also taking place, which can be localised by the peak occurring in the pressure ratio of freely rotating and perfectly ordered systems.~\cite{Mizani-Oettel-Gurin-Varga_SciPostPhysCore_2025,Mizani-Gurin-Varga-Oettel_PhysRevE_2025}. As far as the patchy particles are concerned, they can belong to the second class, since the pair interaction is directional, even if the shape of the particle is spherical. The case of one and two patches on the surface of a hard sphere was investigated in the sticky limit for that special case where the centres of the particles are confined to a straight line~\cite{Lebowitz-Percus_AnnNYAcadSci_1983}. Using the TMM, it was found that orientational ordering is accompanied with dimer formation in the case of a single patch, while in the case of two patches, even long chains and first-order phase transitions can occur~\cite{Lebowitz-Percus_AnnNYAcadSci_1983}. Later, instead of the exact TMM, the well-known Wertheim’s first order thermodynamic theory (TPT1), which is not an exact theory, was applied for some q1D fluids of hard spheres with two patches~\cite{Marshall_JChemPhys_2015}. The comparison of TPT1 and simulation results showed that TPT1 describes quite well the enhanced association and the orientational ordering of the patchy particles in the channel~\cite{Marshall_JChemPhys_2015}. This suggests that TPT1 could perhaps be modified to be exact with suitable extension for inhomogeneous fluids. The recent application of TMM for q1d hard rectangles with two patches at the tips showed that the particles align along the confining channel at low and intermediate densities to maximize the number of bonds, while the bonds between the particles break and the particles take perpendicular directions at very high densities to minimize their length along the channel~\cite{Gurin-Varga_PhysRevE_2022}. Competition between association and orientation ordering effects can lead to a more complex phase behaviour than the two classes mentioned above if the bonding sites are selective and the particle shape is more complicated (disk-like, rod-like or non-convex).

In this work, we investigate some q1D fluids consisting of patchy particles with the goals of 1) extending the class of patchy models that can be solved exactly, 2) boosting new simulation studies, and 3) contributing with exact bulk free energy in the construction of density functionals for anisotropic patchy particles. To achieve this goal, we rely on the TMM, which is tractable if only close neighbours interact, such as first neighbour, second neighbour, etc. interactions are present, in the q1D array of particles. We realise this with constraining the centres of particles to a straight line and making the patchy attraction extremely short ranged. We distribute the patches on anisotropic hard particles and assume that all bonds between patchy particles are located on a line. This occurs when the point of contact of two adjacent particles falls on the confining line. The most common geometric shapes that meet this criterion are the circle and the sphere. Even regular and irregular polygons and polyhedra can satisfy this criterion if the possible orientations of the particle are limited to those angles that result in an additive contact distance.
The consequence of this orientational restriction is that the maximum number of patches is limited to the number of possible orientations. Without specifying the molecular shape and the number of patches, we derive analytical expressions for the Gibbs free energy, the equation of state, order parameter and the average length of the cluster. Using these equations, we devise an orientation dependent free energy formalism, where the density is angle dependent and the resulting mass action law and the association free energy term keeps the format of Wertheim’s association theory of isotopic fluids.
The novelty in our generalised equations is that the association terms are weighted with the orientational distribution of particles.
The minimization of the free energy density with respect to the angle dependent density and the fraction of sites unbounded, provides a set of coupled equations for the equilibrium orientational distribution and the fraction of sites unbounded. We show that choosing pressure as the independent variable instead of density significantly reduces the computational burden, and that the generalised Wertheim theory is identical to the TMM for the q1D sticky patchy particle systems. We do not make an attempt to apply the new theory for specific systems, because the number of possible systems diverge with the number of orientations. The possible applications is left for future studies.
        
\section{Patchy models and distribution functions}

\begin{figure}
  \centering
  \includegraphics[width=1\columnwidth]{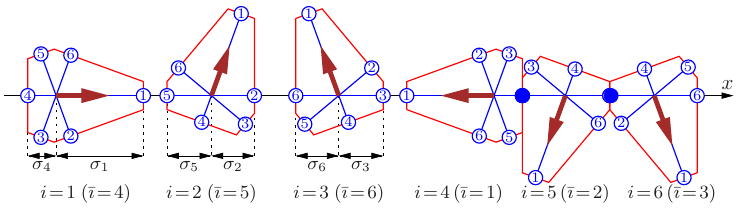}
  \caption{An example of q1D systems of anisotropic sticky hard particles. The particle has 6 possible orientations and bonding sites. The orientational unit vector is denoted by thick brown arrow, which is in a 2D plane in this example but can be arbitrary in 3D in general. The bonding sites are denoted by blue circles. A trimer (3 particles are in bond) is also shown with filled circles of binding pair of sites, numbered with (4,2) and (5,3), respectively. The numbering of orientations and sites is harmonised such a way that in orientation $\o$, sites $\so$ (on the right) and $\sobar$ (on the left) are allowed to form bond. All the relevant information about the shape of the particle bounded by the red curve is contained in the $\sigma_\o$ radii (shown by the blue skeleton), because they determine the contact distance. The intersection of the blue lines is considered as the position of the particle, which is constrained to the $x$ axis.}
\label{fig:particles}
\end{figure}

We study q1D systems of patchy particles, where the interaction between the particles consists of additive hard body repulsion and sticky attraction along the confining line. We assume that a given point of the particle is constrained to a line ($x$-axis) with continuous positional freedom. The position of this point of particle $\p$ is denoted by $x_\p$. Around this point, the particle is allowed to rotate into $\n$ possible orientations that are labeled by $\o$, i.e. $\o \in \{1,\dots,\n\}$. These orientations can be arbitrary in either two- or three dimensions. For a given orientation $\o$, the opposite orientation, when the particle is mirrored to the plane perpendicular to $x$, so the left and right sides are interchanged, is denoted by $\obar$, as can be seen in Fig.~\ref{fig:particles}.

The shape of the hard body is described by orientation dependent radii. In orientation $\o$ the ``right radius'' of a particle is denoted by $\sigma_\o$, while the ``left radius'' is denoted by $\sigma_\obar$, see Fig.~\ref{fig:particles}.  Therefore, the occupied length (diameter) of a particle along the $x$ axis is $d_\o=d_\obar=\sigma_\o+\sigma_\obar$ both in orientations $\o$ and $\obar$. Moreover, the contact distance between a left particle with orientation $\o$ and a right particle with orientation $\ov$ is $\sigma_{\o\ovbar}\equiv\sigma_\o+\sigma_\ovbar$. Therefore, the additive hard body repulsion is defined by the following pair potential,
\begin{equation}
     V_{\o\ov}^\mathrm{HB}(x)
  =  \left\{\begin{array}{lll}
              \infty  & \text{if} & -\sigma_{\ov\obar} < x < \sigma_{\o\ovbar}\\
              0       & 
              \text{otherwise}\hspace{-1cm}
            \end{array}
     \right.
  \label{eq:V^{HB}}
\end{equation}
where $x=x_{\p'}-x_\p$
and $x_\p$ ($x_{\p'}$) are the positions of the particles with orientations $\o$ ($\ov$). 

Besides the hard body interaction, particles are decorated with bonding sites, which can interact selectively with the other sites. The strength of this interaction can even be zero, including the possibility that the particle (or one side of it) behaves as a simple hard body. All orientations are accompanied by two bonding sites along $x$ axis. We use the following convention: for a given orientation $\o$, the site on the right side of the particle is denoted by $\so$ and the site on the opposite side by $\sobar$ as shown in Fig.~\ref{fig:particles}. Note that the roman letters are used for the sticky sites while the italics for the orientations. Because of the short range nature of the interaction, bond can only occur between the bonding sites facing each other. Thus, a particle in the orientation $\o$ can bond with its right neighbour through the site $\so$ and with its left neighbour through the site $\sobar$, and all its other sites are unbonded.

\begin{figure}
  \centering
  \includegraphics[width=0.8\columnwidth]{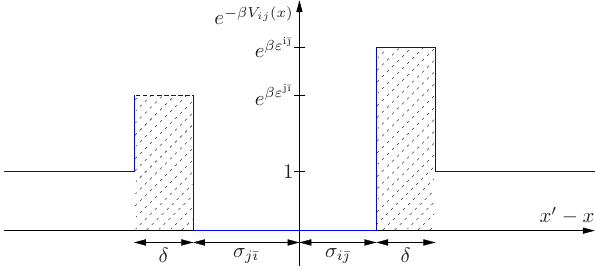}
  \caption{Boltzmann factor of associating hard particles as a function of $x'-x$, where $x$ ($x'$) is the position of the particle with orientation $\o$ ($\ov$). The interaction is hard repulsive in the overlapping region, while the interaction between bonding sites is square-well attraction in the regions with $\delta$ length. The area of the right (left) shaded regions is forced to be $\gamma^{\so\sovbar}$ ($\gamma^{\sov\sobar}$) in the sticky limit: $\delta\rightarrow 0$, $\varepsilon^{\so\sovbar}\rightarrow\infty$, and $\varepsilon^{\sov\sobar}\rightarrow\infty$.}
\label{fig:square-well}
\end{figure}

Now we define the interaction between two patchy particles. To do this, let us consider two anisotropic hard particles in positions $x_\p$ and $x_{\p'}$ and orientations $\o$ and $\ov$, respectively, and calculate $e^{-\beta V_{\o\ov}(x)}$, where $x=x_{\p'}-x_\p$. The pair potential, $V_{\o\ov}(x)$, is the sum of hard body and attractive terms. First, we assume that the bonding sites $\s$ and $\sv$ interact with square-well attraction having $\varepsilon^{\s\sv}$ depth and $\delta$ range. Based on Fig.~\ref{fig:square-well} 
we can write that
\begin{equation}
     e^{-\beta V_{\o\ov}(x)}
  =  \left\{\begin{array}{lll}
        0  & \text{if} & -\sigma_{\ov\obar} < x < \sigma_{\o\ovbar}\\
        e^{\beta\varepsilon^{\sov\sobar}} & \text{if} & -(\sigma_{\ov\obar}+\delta) \leq x \leq -\sigma_{\ov\obar}\\
        e^{\beta\varepsilon^{\so\sovbar}} & \text{if} & \sigma_{\o\ovbar} \leq x\leq \sigma_{\o\ovbar}+\delta\\
        1 & 
        \text{otherwise}.\hspace{-1.5cm}
            \end{array}
     \right.
  \label{eq:V^{SW}}
\end{equation}
In this paper we study only the sticky limit of this attraction potential~\cite{Baxter_JChemPhys_1968}, when the range of the square-wells goes to zero, $\delta\rightarrow 0$, and the depth of them goes to infinity, $\varepsilon^{\s\sv}\rightarrow\infty$, while the condition $\delta e^{\beta \varepsilon^{\s\sv}}=\gamma^{\s\sv}$ is held. Here the symmetric matrix $\gamma^{\s\sv}$ describes the strength of the interaction (relative to the temperature) between the site $\s$ and $\sv$,. Since the square-wells become Dirac delta in the contact points, therefore
\begin{align}
     e^{-\beta V_{\o\ov}(x)}
  =\; &\theta(-x-\sigma_{\ov\obar})
    +\theta(x-\sigma_{\o\ovbar})
 \nonumber \\
  & +\gamma^{\sov\sobar} \delta(x+\sigma_{\ov\obar})
    +\gamma^{\so\sovbar} \delta(x-\sigma_{\o\ovbar}) ,
  \label{Boltzmann-factor}
\end{align}
where $\theta$ is the Heaviside step function.
Now we introduce the bond matrix 
\begin{equation}
\label{Delta_def}
         \Delta_{\o\ov}^{\s\sv}
  \defeq \left\{
          \begin{array}{lll}
             \displaystyle{
               \int_{-\infty}^\infty g^{\mathrm{HB}}(x)(e^{-\beta V_{\o\ov}(x)}-1)\, dx
             } \hspace{-0.9\columnwidth}
             \\
             & \text{if ($\s=\so$ and $\sv=\sovbar$) or ($\s=\sobar$ and $\sv=\sov$)}
             \\~\\
             0 \hspace{5ex} & \text{otherwise}
               \end{array}
     \right.
\end{equation}
where $g^{\mathrm{HB}}(x)$ is the radial distribution function of the 1D hard body fluid which has the same density as the system under study.
In Eq.~(\ref{Delta_def}), as defined above, $\o$ and $\ov$ denote the orientations of the particles, while $\s$ and $\sv$ refer to sites on these particles. Thus, $\Delta_{\o\ov}^{\s\sv}$ describes the interaction of the bond $\s$ of a particle with orientation $\o$ with the bond $\sv$ of an other particle with orientation $\ov$. Note that while $\gamma^{\s\sv}$ describes purely the interaction of sites $\s$ and $\sv$ (therefore it is symmetric), $\Delta_{\o\ov}^{\s\sv}$ also takes into account the steric incompatibility of the sites when the orientations of the particles are fixed. Therefore, in general $\Delta_{\o\ov}^{\s\sv}\neq\Delta_{\o\ov}^{\sv\s}$, however $\Delta_{\o\ov}^{\s\sv}=\Delta_{\ov\o}^{\sv\s}$.
Inserting Eq.~(\ref{Boltzmann-factor}) into Eq.~(\ref{Delta_def}) it is easy to show that for sticky particles
\begin{equation}
\label{Delta}
     \Delta_{\o\ov}^{\s\sv}
  =  \frac{\gamma^{\s\sv}}{1-\eta} 
         (\delta_{\s\so}\delta_{\sv\sovbar}+\delta_{\s\sobar}\delta_{\sv\sov}) ,
\end{equation}
where we used that $g^{\mathrm{HB}}(\sigma_{kl})=\frac{1}{1-\eta}$ with $\eta$ being the 1D packing fraction~\cite{Salsburg-Zwanzig-Kirkwood_JCP_1953}.

Besides the interaction between the particles we take into account that particles can interact with a uniform external field, that is a particle with orientation $\o$ has external energy $U_\o$.

Now we introduce some bulk and structural quantities.
Let $N_\o$ be the number of particles with orientation $\o$, ($\o \in \{1,\dots,\n\}$), and $N=\mysum{\o} N_\o$ is the total number of particles.
$N_\o^{\uu\s}$ is similar to $N_\o$ with the condition that the site $\s$ of the particle is unbonded.
$N_{\o\ov}$ is the number of nearest neighbour pairs, where the left particle has $\o$ and the right one has $\ov$ orientations.
$N_{\o\ov}^\uu$ is the same but the pair is unbonded, i.e. $x>\sigma_{\o\ovbar}$.
The density is $\rho=N/L$ and $\rho_\o=\langle N_\o\rangle /L$ where $L$ is the length of the system and the $\langle\cdot\rangle$ denotes the ensemble average.
Then the 1D packing fraction is given by $\eta=\mysum{\o} \rho_{\o} d_{\o}$.
We can define the normalized orientational distribution function (ODF) as
\begin{equation}
     f_\o
  \defeq \frac{\langle N_\o \rangle}{N}
  =  \frac{\rho_\o}{\rho} ,
  \label{eq:f}
\end{equation}
which is, in other words, the probability of finding a given particle (e.g. particle 1) in the orientation $\o$. 
With the help of the partial nearest neighbour pair distribution function, 
$f^{(2)}_{\o\ov}(x)$, (which is the probability density distribution of finding a nearest neighbour pair with distance $x$ where the left particle's orientation is $\o$ and right particle's orientation is $\ov$) we can define some useful distribution functions as follows
\begin{equation}
     f^{(2)}_{\o\ov}
  \defeq \frac{\langle N_{\o\ov} \rangle}{N} = \int_0^\infty f^{(2)}_{\o\ov}(x)\, dx ,
  \label{eq:f2_ij}
\end{equation}
\begin{equation}
     f^{(2)\uu}_{\o\ov}
  \defeq \frac{\langle N_{\o\ov}^\uu \rangle}{N}
  =  \lim_{\delta\rightarrow 0}
     \int_{\sigma_{\o\ovbar}+\delta}^\infty f^{(2)}_{\o\ov}(x)\, dx ,
  \label{eq:f2u}
\end{equation}
and
\begin{equation}
     f_\o^{\uu\s}
  \defeq \frac{\langle N_\o^{\uu\s} \rangle}{N}
  =  \left\{\begin{array}{ll}
              \mysum{\ov} f^{(2)\uu}_{\o\ov} & \text{if\ }  \s=\so\\
              \mysum{\ov} f^{(2)\uu}_{\ov\o}  & \text{if\ }  \s=\sobar\\
              \langle N_\o \rangle/N  & \text{otherwise}. \\
            \end{array}
     \right.
  \label{eq:fu}
\end{equation}
Moreover, the total nearest neighbour pair distribution function is given by
\begin{equation}
     f^{(2)}(x)
  \defeq \mysum{\o\ov} f^{(2)}_{\o\ov} .
  \label{eq:f2}
\end{equation}
It is also important to define the key quantity of Wertheim's association theory~\cite{Wertheim_JStatPhys_1984,Wertheim.1_JStatPhys_1984,Wertheim.2_JStatPhys_1984,Wertheim.3_JStatPhys_1986,Wertheim.4_JStatPhys_1986,Wertheim_JChemPhys_1987}
\begin{equation}
     \x_\o^{\u\s}
  \defeq \frac{\langle N_\o^{\uu\s} \rangle}{\langle N_\o \rangle} ,
  \label{eq:xu}   
\end{equation}
which is the fraction of particles with unbonded site $\s$ among the particles with orientation $\o$. Because a particle with orientation $\o$ can interact only with its right and left neighbours through its sites $\so$ and $\sobar$, while all other sites of it are unbonded, therefore $N_\o^{\uu\s} = N_\o$ and $\x_\o^{\u\s} = 1$ if $\s\notin\{ \so,\sobar \}$.

If two, three or more particles are in contact, they form dimers, trimers etc. molecules. We denote the total number of molecules by ${\cal N}$, and $\rho^{\mathrm{mol}}=\langle{\cal N}\rangle/L$ is the molecule density.
Every molecule (chain of particles) has a single left (and right) end which is a particle with an unbounded left (right) site, therefore one can show that
\begin{equation}
  \rho^{\mathrm{mol}} =
     \mysum{\o} \rho_\o \x_\o^{\u\so}
  =  \mysum{\o} \rho_\o \x_\o^{\u\sobar} .
  \label{N_mol}
\end{equation}

\section{Transfer matrix method}

One dimensional models of the kind we consider here can be solved exactly using the transfer matrix formalism, which is briefly summarized below.
In the calculation of isobaric partition function, $Z(N\!,P,T)$, imposing periodic boundary conditions such that $x_{N+1}\equiv x_1$ and $L = \sum_{\p=1}^{N} x_{\p,\p+1}$, where $x_{\p,\p+1}=x_{\p+1}-x_\p$ we obtain that
\begin{align}
\label{Z_def}
     Z(N\!,P,T)
 &=  \int\limits_0^\infty\dd x_{1,2} \cdots \int\limits_0^\infty\dd x_{N,1}
     \mysum{\o_1} \cdots \mysum{\o_N}
     \nonumber \\
 &\mbox{}\hspace{6ex}
     e^{-\beta\sum_{\p=1}^N\left( V_{\o_{\p}\o_{\p+1}}(x_{\p,\p+1})
       +   U_{\o_n}
       + P x_{\p,\p+1} \right)} \nonumber\\
 &=  \mysum{\o_1} \cdots \mysum{\o_N}
     K_{\o_1\o_{2}} \cdots K_{\o_N\o_1}
  =  \Tr K^N .
\end{align}
Note that the $1/N!$ prefactor is discarded due to $x_1\leq x_2\leq\dots\leq x_N$ condition is satisfied.
In Eq.~(\ref{Z_def}) $P$ is the pressure,  the summation index $\o_{\p} = 1,\dots ,\n$ is the orientation of particle $\p$ and the transfer matrix, $K$, is defined by
\begin{align}
\label{K}
     K_{\o\ov}
 &=  \int\limits_0^\infty\dd x\, e^{-\beta( V_{\o\ov}(x) +U_i + P x )}
 \nonumber \\
 &=  \frac{e^{-\beta (U_i+P \sigma_{\o\ovbar}})}{\beta P}
     \left( 1 + \beta P \gamma^{\so\sovbar} \right) ,
\end{align}
where in the second equality we used Eq.~(\ref{Boltzmann-factor}).

In the case of the models we study, $K$ is not necessarily a symmetric matrix, therefore in general there is no orthonormalised basis formed by the eigenvectors of $K$, i.e. the eigenvectors are usually not orthogonal to each other, moreover, the number of the eigenvectors can be less than the number of the rows of $K$. However, all elements of $K$ are positive, thus the Perron-Frobenius theorem~\cite{Cuesta-Sanchez_JStatPhys_2004} guarantees that $K$ has at least one right and one left eigenvector, they are denoted by $\psi_{\nul}^R$ and $\psi_{\nul}^L$, respectively, and the corresponding eigenvalue, which is the so called dominant  eigenvalue (because any other eigenvalue of $K$ are strictly smaller in absolute value) is denoted by $\lambda_{\nul}$. They satisfy the following left and right hand side eigenvalue equations,
\begin{subequations}\begin{align}
  \mysum{\ov} \psi^L_\ov K_{\ov\o} &= \lambda \psi^L_\o , \label{eig.eqs-L} \\
  \mysum{\ov} K_{\o\ov} \psi^R_\ov &= \lambda \psi^R_\o . \label{eig.eqs-R}
\end{align}  \label{eig.eqs}\end{subequations}
All the coordinates of these eigenvectors are positive, and the absolute values of all the other eigenvalues of $K$ are necessarily strictly less than $\lambda_{\nul}$. The eigenvectors are determined up to an arbitrary constant, but as a normalization convention, we fix that $\mysum{\o} \psi^R_\o \psi^L_\o = 1$.

It also follows that the one-dimensional subspace generated by $\psi_{\nul}^R$ (denoted by $V_0={\cal L}(\psi_{\nul}^R)$) is an invariant subspace of $K$, and the orthogonal complement of the one-dimensional subspace generated by $\psi_{\nul}^L$ (denoted by $V_1=({\cal L}(\psi_{\nul}^R))^\perp$) is also an invariant subspace of $K$. Denoting the projection along $V_1$ onto $V_0$ by $P_0$ and the projection along $V_0$ onto $V_1$ by $P_1$, it is clear that $P_0+P_1=1$ and $P_0P_1=P_1P_0=0$. Furthermore, $K=K_0+K_1$, where $K_0\defeq P_0 K P_0 = \lambda_{\nul} P_0$ and $K_1 \defeq P_1 K P_1$. Since all eigenvalues are smaller than $\lambda_{\nul}$, it can be proved, based on the Jordan normal form of the matrix, that $|\langle\varphi| K_1|\varphi'\rangle| <  c\lambda_{\nul}|\langle\varphi|\varphi'\rangle|$ for any vector $\varphi$ and $\varphi'$, where $0<c<1$. It also follows that
\begin{equation}
  \label{decomp}
     |\langle\varphi| K_1^N|\varphi'\rangle|
  <  c^N\lambda_{\nul}^N|\langle\varphi|\varphi'\rangle|
  \quad \text{where\ } |c|<1 .
\end{equation}
  This inequality allows us to extend the transfer matrix formalism to non-symmetric matrices, because $Z$ can be written as 
\begin{equation}
\label{Z_no.T.lim}
     Z(N\!,P,T)
  =  \Tr K_0^N + \Tr K_1^N 
  = \lambda_{\nul}^N \left( 1 + \frac{\Tr K_1^N }{\lambda_{\nul}^N} \right) .
\end{equation}
In thermodynamic limit, when $N\rightarrow\infty$, the second term vanishes because of Eq.~(\ref{decomp}), therefore
\begin{equation}
\label{Z}
     Z(N\!,P,T)
  =  \lambda_{\nul}^N .
\end{equation}
Note that the Gibbs free energy can be calculated with $\lambda_{\nul}$, because $\beta G = -\ln Z(N\!,P,T)$. As $G/N$ corresponds to the chemical potential, $\mu$, we get that $\lambda_{\nul}=e^{-\beta\mu}$.

We mention, that in the absence of external field there is a simple relation between the right and left eigenvectors, namely $\psi^L_{\nul\o} = \psi^R_{\nul\obar}$. This equation can be easily proved based on the left and right eigenvalue equations, Eqs.~(\ref{eig.eqs}), and using that the reflection on the $y$ axis does not change the energy of a nearest-neighbour pair of particles when the external field is zero, therefore $K_{\o\ov}=K_{\ovbar\obar}$. But this is not true in general, when $U_\o \neq U_\obar$.

In the transfer matrix formalism, the orientation distribution function can be computed in a similar way as the partition function, because the integrals are factorized and can again be expressed using the transfer matrix, and we obtain
\begin{align}
  \label{eq:f_j_trmx}
     f_\o
  =  \frac{1}{Z}&
     \int\limits_0^\infty\dd x_{1,2} \cdots \int\limits_0^\infty\dd x_{N,1}
     \mysum{\o_1} \cdots \mysum{\o_N}  \delta_{\o\o_1} \times
     \nonumber\\&\mbox{}\hspace{3ex}
     e^{-\beta\sum_{\p=1}^N\left( V_{\o_{\p}\o_{\p+1}}(x_{\p,\p+1}) + U_{\o_{\p}} + P x_{\p,\p+1} \right)}
     \nonumber\\
  =  \frac{1}{Z}&
     \mysum{\o_2} \cdots \mysum{\o_N}
     K_{\o\o_2}  K_{\o_2\o_3}\cdots K_{\o_{N-1}\o_N} K_{\o_N\o}
  =  \frac{(K^N)_{\o\o}}{Z} .
\end{align}
Hence, in thermodynamic limit, based on Eq.~(\ref{decomp}) and that $K^N=K_0^N+K_1^N$, similarly as Eq.~(\ref{Z}) was obtained from  Eq.~(\ref{Z_no.T.lim}), we conclude that
\begin{align}
  \label{f_j-trmx}
     f_\o
  =  \psi^L_\o \psi^R_\o .
\end{align}
In a similar way, the probability density, which defined for $x>0$ as $ f^{(2)}_{\o\ov}(x) \defeq \langle \delta(x-x_{1,2})\delta_{\o_1\o}\delta_{\o_2\ov} \rangle$, that is
\begin{align}
\label{eq:f_def}
     f^{(2)}_{\o\ov}&(x) 
  =  \frac{1}{Z}
     \int\limits_0^\infty\dd x_{1,2} \cdots \int\limits_0^\infty\dd x_{N,1}
     \mysum{\o_1} \cdots \mysum{\o_N} \delta(x-x_{1,2}) \times
     \nonumber\\ &
     \delta_{\o_1\o}\delta_{\o_2\ov}
     e^{-\beta\sum_{\p=1}^N\left( V_{\o_{\p}\o_{\p+1}}(x_{\p,\p+1}) + U_{\o_{\p}} + P x_{\p,\p+1} \right)}
\end{align}
can be calculated in the thermodynamic limit as follows~\cite{Gurin-Varga_JCP_2013}
\begin{align}
     f^{(2)}_{\o\ov}(x)
 &=  \frac{1}{Z}
     \mysum{\o_3} \cdots \mysum{\o_N} K_{\ov\o_3}  K_{\o_3\o_4}\cdots\nonumber\\
    & \mbox{}\hspace{10ex}
     \cdots K_{\o_{N-1}\o_N} K_{\o_N\o}
     e^{-\beta\left( V_{\o\ov}(x) + U_{\o} + P x \right)} \nonumber\\
 &=  \frac{\psi^L_\o \psi^R_\ov}{\lambda_{\nul}} e^{-\beta\left( V_{\o\ov}(x) + U_{\o} + P x \right)} ,
\end{align}
and using Eq.(\ref{Boltzmann-factor}), taking into account that $x>0$, the nearest neighbour pair distribution function is 
\begin{align}
     f^{(2)}_{\o\ov}(x) 
  =  \frac{\psi^L_\o \psi^R_\ov}{\lambda_{\nul}}
     \left[ \theta(x-\sigma_{\o\ovbar})
           +\gamma^{\so\sovbar} \delta(x-\sigma_{\o\ovbar})
     \right] e^{-\beta(U_{\o} + P x)} .
  \label{eq:f2(x)_trmx}
\end{align}
From Eqs.~(\ref{eq:f2_ij}) and (\ref{eq:f2(x)_trmx}) immediately follows that
\begin{equation}
     f^{(2)}_{\o\ov}
  =  \frac{\psi^L_\o \psi^R_\ov}{\lambda_{\nul}}
     \frac{e^{-\beta(U_{\o} + P \sigma_{\o\ovbar})}}{\beta P}
     \left( 1 + \beta P \gamma^{\so\sovbar} \right) ,
  \label{eq:f2_trmx}
\end{equation}
and from Eqs.~(\ref{eq:f2u}) and (\ref{eq:f2(x)_trmx}) 
\begin{equation}
     f^{(2)\uu}_{\o\ov}
  =  \frac{\psi^L_\o \psi^R_\ov}{\lambda_{\nul}}
     \frac{e^{-\beta(U_{\o} + P \sigma_{\o\ovbar})}}{\beta P} .
  \label{eq:f2u_trmx}
\end{equation}

Using the above distribution functions, we can get all important quantities. For example, in the transfer matrix formalism the equation of state generally obtained from the standard thermodynamic relation between the Gibbs free energy and the pressure, which leads to the formula $\rho^{-1}=-\frac{\partial\ln\lambda_{\nul}}{\partial \beta P}$. However, based on the nearest-neighbour distribution functions derived above, there is an alternative way, since the average distance between two neighbouring particles, $\langle x \rangle$, is the inverse of the density, thus
\begin{equation}
     \frac{1}{\rho}
  =  \langle x \rangle
  =  \int x f^{(2)}(x) dx
  =  \mysum{\o\ov} \int x f^{(2)}_{\o\ov}(x) dx .
  \label{eq:<x>}
\end{equation}
 
\section{Generalised Wertheim's TPT1 theory}

The well-known Wertheim's TPT1~\cite{Wertheim_JStatPhys_1984,Wertheim.1_JStatPhys_1984,Wertheim.2_JStatPhys_1984,Wertheim.3_JStatPhys_1986,Wertheim.4_JStatPhys_1986,Wertheim_JChemPhys_1987} seems to be the most natural choice for the study of q1D associating liquids, as it has proven to be very successful in determining the phase equilibrium, structure, size distribution and equilibrium constant of hydrogen-bonded molecular liquids~\cite{Jackson-Chapman-Gubbins.1_MolPhys_1988,PARICAUD200287,10.1063/5.0098882} and patchy colloids~\cite{Teixeira-Tavares_CurrOpinColloidInterfaceSci_2017}. Its application is also justified by its incorporation into density functional theories to describe the inhomogeneous structure of network forming associating particles in confinement~\cite{10.1063/5.0180795,10.1063/1.4776759,D2RA02162E}, and its generalization to dimerizing and hydrogen-bonding anisotropic particles forming liquid crystal phases~\cite{10.1063/1.473693}. However, we cannot afford to use any of the existing TPT1 versions because, unlike TMM, they are only perturbative theories. Here we devise a new formalism, which provides the exact thermodynamic properties of q1D fluids of anisotropic patchy particles in the sticky limit.

To be as close as possible to the concept of Wertheim~\cite{Wertheim_JStatPhys_1984,Wertheim.1_JStatPhys_1984,Wertheim.2_JStatPhys_1984,Wertheim.3_JStatPhys_1986,Wertheim.4_JStatPhys_1986,Wertheim_JChemPhys_1987}, we work in the canonical ensemble, and we start from a free energy equation, which can describe the phase behaviour of $\n$ component mixture having $\n$ associating sites in the presence of homogeneous external field, which acts on all particles. Now the index $\o$ represents the component $\o$, which will be the orientation of the particle in the end. We also use the fraction of a site not bonded, which is the key quantity of the Wertheim theory. The input of our starting equation are the number densities, $\rho_{\o}$, and the bond matrix, $\Delta_{\o\ov}^{\s\sv}$, see Eqs.~(\ref{Delta_def}--\ref{Delta}), while the fraction of a site $\s$ not bonded on component $\o$, that is $X_{\o}^{\s}$ is the output of the theory. We assume that the free energy density is given by
\begin{align}
     \frac{\beta F}{L}
 = & \mysum\o \rho_\o (\ln \rho_\o-1) -\rho\ln(1-\eta)
 \nonumber \\ &   
    +\mysum{\o\s} \rho_\o(\ln \x_\o^{\u\s}-\x_\o^{\u\s}+1)
    -\frac{1}{2} \mysum{\o\ov\s\sv}
       \rho_\o \rho_\ov \x_\o^{\u\s} \x_\ov^{\u\sv} \Delta _{\o\ov}^{\s\sv}
\nonumber \\ &
    +\mysum\o \rho_{\o} \beta U_{\o} .
  \label{W:free_energy_density}
\end{align}
In Eq.~(\ref{W:free_energy_density}) the first term is the ideal gas part, the second one is the hard body exclusion part, the third and the fourth terms together take into account the association and the last term is the external field contribution. The ideal and the external field contributions are known to be are exact in Eq.~(\ref{W:free_energy_density}). The hard body contribution is exact only for additive mixtures, but not for nonadditive ones~\cite{%
  Lebowitz-Zomick_JChemPhys_1971,Schmidt_PRE_2007
}. Regarding the association terms, we will show later that it is also exact if all bonding sites are sticky.

In order to obtain the equilibrium properties of the associating fluid, the free energy density, Eq.~(\ref{W:free_energy_density}), must be minimised with respect to $\x_\o^\s$, i.e.
\begin{equation}
  \frac{\delta(\beta F/L)}{\delta X_{\o}^{\s}}=0 .
 \label{conditions_of_equilibrium:X---Szabi}
\end{equation}
After the substitution of Eq.~(\ref{W:free_energy_density}) into Eq.~(\ref{conditions_of_equilibrium:X---Szabi}) and using the $\Delta_{\o\ov}^{\s\sv}=\Delta_{\ov\o}^{\sv\s}$ symmetry property, we end up with the well-known mass action law for $X_{\o}^{\s}$, which is given by
\begin{equation}
  \frac{1}{\x_\o^{\u\s}}=1+\mysum{\ov\sv} \rho_\ov \x_\ov^{\u\sv} \Delta_{\o\ov}^{\s\sv} .
  \label{W:mass_action_law}
\end{equation}
In principle, this equation corresponds to $\n^{2}$ coupled equations, because $\o,\s\in\{1, \ldots, m\}$. However, there are only $2\n$ non-trivial equations, because only two sites of all components can be bonded, while the other sites are never bonded. Eq.~(\ref{W:mass_action_law}) shows that $X_{\o}^{\s}=1$ if the elements of $\Delta_{\o\ov}^{\s\sv}$ appearing in Eq.~(\ref{W:mass_action_law}) are all zero. It can be also seen from Eq.~(\ref{W:mass_action_law}) that if $\Delta_{\o\ov}^{\s\sv}>0$, Eq.~(\ref{W:mass_action_law}) provides physically meaningful results for $X_{\o}^{\s}$, because $X_{\o}^{\s}$ is forced to be between 0 and 1. Therefore $2\n$ solutions of Eq.~(\ref{W:mass_action_law}) are $X_{\o}^{\o}$ and $X_{\o}^{\obar}$, ($\o\in\{1,\dots,\n\}$), which satisfy $0<X_{\o}^{\o}, X_{\o}^{\obar} \leq 1$ condition, while $X_{\o}^{\s}=1$ for $\s\notin\{ \so,\sobar \}$. Our free energy expression, Eq.~(\ref{W:free_energy_density}), can be rewritten into the well-known form derived by Wertheim~\cite{Wertheim_JStatPhys_1984,Wertheim.1_JStatPhys_1984,Wertheim.2_JStatPhys_1984,Wertheim.3_JStatPhys_1986,Wertheim.4_JStatPhys_1986,Wertheim_JChemPhys_1987}, if the bond matrix term of Eq.~(\ref{W:free_energy_density}) is expressed with the help of Eq.~(\ref{W:mass_action_law}). After the combination of Eqs.~(\ref{W:free_energy_density}) and (\ref{W:mass_action_law}), we get that
\begin{align}
  \frac{\beta F}{L}=& \mysum{\o} \rho_{i} (\ln \rho_{i}-1)-\rho \ln (1-\eta)
\nonumber \\ &
  + \mysum{\o\s} \rho_{i}\left( \ln \x_\o^{\u\s}-\frac{\x_\o^{\u\s}}{2}+\frac{1}{2}\right)+\mysum\o \rho_{\o} \beta U_{\o} \,,
  \label{Wertheim_free_energy_density}
\end{align}
which is shorter than Eq.~(\ref{W:free_energy_density}) and does not contain $\Delta_{\o\ov}^{\s\sv}$. Moreover, it can be seen clearly that if $X_{\o}^{\s}=1$ (the case of no bond) the association term of the free energy is zero, see the third term of Eq.~(\ref{Wertheim_free_energy_density}).

Now we continue with the chemical potential of component $\o$ of the mixture, which can be obtained either from Eq.~(\ref{W:free_energy_density}) or Eq.~(\ref{Wertheim_free_energy_density}) as $\mu_{\o}=\left.\frac{\partial F}{\partial N_{\o}}\right|_{T, L, N_{\ov \neq \o}}$. As $X_{\o}^{\s}$ depends on the component densities, it can be shown that $\beta \mu_{\o}=\frac{\partial(\beta F/L)}{\partial \rho_{\o}}+\mysum{\o} \frac{\delta(\beta F/L)}{\delta X_{\o}^{\s}} \frac{\partial X_{\o}^{\s}}{\partial \rho_{\o}}$. This equation shows that Eq.~(\ref{W:free_energy_density}) is more practical because together with Eq.~(\ref{conditions_of_equilibrium:X---Szabi}), only the explicit dependence of Eq.~(\ref{W:free_energy_density}) from $\rho_{\o}$ contributes to $\beta \mu_{\o}$. We get that
\begin{align}
  \beta \mu_{\o}=&\ln \rho_\o-\ln (1-\eta)+\frac{\rho d_\o}{1-\eta}+\mysum{\s} \ln \x_\o^{\u\s}
   \nonumber \\ &
   -\frac{d_\o}{2(1-\eta)} \mysum{\ov\s} \rho_\ov\left(1-\x_\ov^{\u\s}\right)
   + \beta U_{\o}  \,,
  \label{W:mu(rho,X)}
\end{align}
Note that the use of mass action law, Eq.~(\ref{W:mass_action_law}), made it possible to get rid of $\Delta_{\o\ov}^{\s\sv}$ in $\beta \mu_{\o}$. We can also express the pressure of the system using the relationship between the free energy, the chemical potentials and the pressure as $\beta P=\frac{-\beta F}{L}+\mysum{\o} \rho_{\o}\beta\mu_{\o}$. Using Eqs.~(\ref{Wertheim_free_energy_density}), (\ref{W:mu(rho,X)}) and (\ref{eq:f}) we find that
\begin{equation}
  \beta P=\frac{\rho-\frac{1}{2} \mysum{\o\s} \rho_\o\left(1-\x_\o^{\u\s}\right)}{1-\eta}
  =\frac{\rho}{1-\eta} \mysum{\o} f_\o\x_\o^{\u\so} \,.
  \label{W:P}
\end{equation}

We can see from Eqs.~(\ref{Wertheim_free_energy_density}), (\ref{W:mu(rho,X)}) and (\ref{W:P}) that the bond matrix, $\Delta_{\o\ov}^{\s\sv}$, does not have explicit contribution to the thermodynamics of 1D associating mixtures, but it has implicit effect through the fractions of sites not bonded, see Eq.~(\ref{W:mass_action_law}). Therefore the phase behaviour of $\n$ component 1D mixture with the input of component densities $\rho_{1}, \ldots, \rho_{\n}$ (or equivalently total density and mole fractions) can be examined with the solution of Eq.~(\ref{W:mass_action_law}) for $X_{\o}^{\s}$ to get the free energy, the chemical potentials and the pressure from Eqs.~(\ref{Wertheim_free_energy_density}--\ref{W:P}), which is the standard procedure in the Wertheim theory.

The reason for choosing $m$ bonding sites for all particles in the $\n$ component mixture in the above description was that the mixture can be viewed as an orientationally frozen state of a one component fluid, where the particles have $\n$ different orientations in the channel. Therefore if a particle is frozen in the orientation $\o$, then it belongs to component $\o$, but it still has $\n$ bonding sites. Now we show that the above theory can be applied for the one component fluid of sticky and anisotropic hard bodies having $\n$ different orientational states. If the particle can rotate into $\n$ different orientational states, the chemical potential of all states must be the same, i.e. our new condition is that $\beta \mu=\beta \mu_{1}=\beta \mu_{2}=\ldots=\beta \mu_{m}$. Using this condition and Eq.~(\ref{W:mu(rho,X)}), we can express the mole fraction of component $i$, which corresponds to the orientational distribution function, as follows
\begin{equation}
  f_{\o}=\frac{1}{c} \frac{e^{-\beta U_{\o}-\frac{\rho d_\o}{1-\eta}\left(1-\frac{1}{2}\mysum{\ov\s} f_{\ov}\left(1-\x_\ov^{\u\s}\right)\right)}}{\x_\o^{\u\so}\x_\o^{\u\sobar}} ,
  \label{W:f}
\end{equation}
where we used that $\myprod{\s} \x_\o^{\u\s}=\x_\o^{\u\so} \x_\o^{\u\sobar}$, because $\x_\o^{\u\s} = 1$ if $\s\notin\{ \so,\sobar \}$. The normalisation constant $c$ can be obtained from $\mysum{\o} f_{\o}=1$. We can see that Eq.~(\ref{W:f}) corresponds to $\n$ coupled equations involving the variables $f_{\o}$ and  $\x_\o^{\u\s}$. As $\rho_{\o}=\rho f_{\o}$, we can express all quantities ($\eta, X_{\o}^{\s}, \beta F/L, \beta\mu$ and $\beta P$) as a function of $\rho$. Therefore, the procedure to obtain the equilibrium properties of the one component system with $m$ orientational states using the density $\rho$ and bond matrix $\Delta_{\o\ov}^{\s\sv}$ as input is to solve the set of equations of Eq.~(\ref{W:mass_action_law}) and Eq.~(\ref{W:f}) to get $X_{\o}^{\s}$ and $f_{\o}$. In the most general case the number of equations is $3\n$ , because Eq.~(\ref{W:mass_action_law}) corresponds to $2\n$ equations, while Eq.~(\ref{W:f}) to $\n$ equation. Having obtained the equilibrium $X_{\o}^{\s}$ and $f_{\o}$, we get $\beta F / L, \beta \mu$ and $\beta P$ from Eqs.~(\ref{Wertheim_free_energy_density}--\ref{W:P}). In the next section we show that it is worthwhile to choose pressure as the independent variable instead of density to reduce the computational burden and to prove that our formalism is exact.

\section{The equivalence of TMM and generalised TPT1}

Our aim now is to compare the transfer matrix method and the generalised TPT1 theory and to show that the results are the same, i.e. the two theories are in fact equivalent.

There are two important differences between the two theories. On the one hand, our generalised TPT1 theory is formulated on the canonical $(N\!, L, \!T)$ ensemble, while the TMM is on the isobaric $(N\!, P, T)$ ensemble. Therefore, in the spirit of the TMM, we rewrite the equations of the generalised TPT1 to consider pressure as an independent variable. We will see that it helps to simplify the working equations (Eqs.~(\ref{W:mass_action_law}) and (\ref{W:f})). Thus the important message of TMM is that the isobaric ensemble is more convenient than the canonical one, because equations are simpler.
On the other hand, in the formalism of Wertheim theory, the fraction of sites unbounded, $\x_\o^{\u\s}$, play an important role. Therefore we will reformulate the results of the TMM in the language of $\x_\o^{\u\s}$. In this way, the equations of the TMM give a new look, a new message, and at the same time help to prove the equivalence of the two formalisms. Although the TMM and the generalised TPT1 theory are defined in different ensembles, we emphasize here that we study only bulk properties, where the ensembles are equivalent.

First we carry out the reformulation of the equations of the generalised TPT1 using the pressure.
The comparison of Eqs.~(\ref{W:P}) and (\ref{W:f}) shows that the ODF can be rewritten in the following simpler form,
\begin{equation}
  f_\o
  =\frac{1}{c} \frac{e^{-\beta(U_{\o} + P d_\o)} }{\x_\o^{\u\so} \x_\o^{\u\sobar}} ,
  \quad\text{where}\quad
  c={\mysum{\o} \frac{e^{-\beta(U_{\o} + P d_\o)}}{\x_\o^{\u\so} \x_\o^{\u\sobar}}} .
  \label{W:f_i}
\end{equation}
This equation shows that we can determine the ODF if $\x_\o^{\u\s}$ is known, that is, the fraction of sites unbounded plays the central role in the Wertheim theory.
It is important to point out here, that we do not reformulate the generalised TPT1 theory in the language of the  $(N\!, P, T)$ ensemble, i.e. we do not define a Gibbs free energy instead of the Helmholtz free energy given by Eq.~(\ref{W:free_energy_density}). We use only the relation between density, packing fraction and pressure, Eq.~(\ref{W:P}), which is in fact the equation of state, to express $\rho/(1-\eta)$ as a function of $P$.
Technically, we can say the following. The generalised TPT1 theory is a perturbation theory that describes the behaviour of a system relative to a reference hard body system. The radial distribution function of the reference system, $g^{\mathrm{HB}}(x)$, is explicitly included in the definition of the bond matrix, $\Delta_{\o\ov}^{\s\sv}$, see Eq.~(\ref{Delta_def}). Although $g^{\mathrm{HB}}(x)$ depends on the density of the hard-body system, we emphasize here that if we switch viewpoints and choose pressure as the independent variable, then $g^{\mathrm{HB}}(x)$ is the radial distribution function of the hard-body system whose density, but not pressure, is the same as that of the patchy system under study.

To see the connection with the TMM, we should express the ODF with  $\x_\o^{\u\s}$, instead of $\psi_\o$. For this reason, from Eq.~(\ref{f_j-trmx}) and (\ref{eq:f2u_trmx}) we get
\begin{align}
     \x_\o^{\u\so}
  =  \frac{\mysum\ov N_{\o\ov}^\uu}{N_\o}
  =  \frac{\mysum\ov f_{\o\ov}^{(2)\uu}}{f_\o}
  =  \mysum\ov \frac{\psi^R_\ov}{\psi^R_\o} \frac{1}{\lambda_{\nul}}
             \frac{e^{-\beta(U_{\o} + P \sigma_{\o\ovbar})}}{\beta P} ,
  \label{eq:x_s^us}
\end{align}
and similarly,
\begin{align}
     \x_\o^{\u\sobar}
  =  \frac{\mysum\ov N_{\ov\o}^\uu}{N_\o}
  =  \mysum\ov \frac{\psi^L_\ov}{\psi^L_\o} \frac{1}{\lambda_{\nul}}
             \frac{e^{-\beta(U_{\ov} + P \sigma_{\ov\obar})}}{\beta P} .
  \label{eq:x_s^usbar}
\end{align}
It is important to note that from Eqs.~(\ref{eq:x_s^us}) and (\ref{eq:x_s^usbar}) follows that $\x_\o^{\u\sobar}=\x_\obar^{\u\sobar}$ when the external field is zero. Using the additivity of the hard body repulsion, $\sigma_{\o\ovbar}=\sigma_\o+\sigma_{\ovbar}$, it follows from Eq.~(\ref{eq:x_s^us}) that
\begin{align}
     \frac{\x_\o^{\u\so}}{\x_\ov^{\u\sov}} e^{-\beta P(\sigma_\ov-\sigma_\o)}
     \frac{e^{-\beta U_\ov}}{e^{-\beta U_\o}}
  =  \frac{\psi^R_\ov}{\psi^R_\o} ,
  \label{eq:x_s^us/x_sv^usv}
\end{align}
and on the same way from Eq.~(\ref{eq:x_s^usbar}) that
\begin{align}
     \frac{\x_\o^{\u\sobar}}{\x_\ov^{\u\sovbar}} e^{-\beta P(\sigma_\ovbar-\sigma_\obar)}
  =  \frac{\psi^L_\ov}{\psi^L_\o} .
  \label{eq:x_s^usbar/x_sv^uvbar}
\end{align}
Taking the product of the above two equations, using Eq.~(\ref{f_j-trmx}), furthermore, by summing over the values of the index $\ov$, and using that the ODF is normalized, we obtain Eq.~(\ref{W:f_i}). Thus, we have proved that the ODF coming from the generalised TPT1 theory, Eq.~(\ref{W:f_i}) and so Eq.~(\ref{W:f}) is exact.

In Eq.~(\ref{W:f_i}) the ODF at a given pressure is explicitly expressed in terms of $\x_\o^{\u\s}$. It shows that using the pressure instead of the density, the coupled equations of Eqs.~(\ref{W:mass_action_law}) and (\ref{W:f}) can be decoupled, and the mass action law should be written in terms of $\x_\o^{\u\s}$ and $P$ only, without $f_{\o}$. For this reason, we substitute $\Delta_{\o\ov}^{\s\sv}$ from Eq.~(\ref{Delta}) into Eq.~(\ref{W:mass_action_law}), then substitute the factor $\rho/(1-\eta)$ expressed as a function of $P$ from Eq.~(\ref{W:P}), and we get that
\begin{equation}
  \frac{1}{\x_\o^{\u\s}}=1+\beta P \frac{\mysum{\ov\sv} f_\ov \x_\ov^{\u\sv} \gamma^{\s\sv}(\delta_{\s\so}\delta_{\sv\sovbar}+\delta_{\s\sobar}\delta_{\sv\sov})}{\mysum{\ov} f_\ov \x_\ov^{\u\sov}} .
  \label{W:mass_action_law_final}
\end{equation}
Finally the ODF dependence of Eq.~(\ref{W:mass_action_law_final}) can be eliminated using  Eq.~(\ref{W:f_i}) as follows
\begin{equation}
  \frac{1}{\x_\o^{\u\s}}=1+\beta P \frac{\mysum{\ov\sv} \frac{e^{-\beta(U_\ov + P d_\ov)}}{\x_\ov^{\u\sovbar}} \gamma^{\s\sv}(\delta_{\s\so}\delta_{\sv\sovbar}+\delta_{\s\sobar}\delta_{\sv\sov}) }{\mysum{\ov} \frac{e^{-\beta(U_\ov + P d_\ov)} }{ \x_\ov^{\u\sovbar}}} .
  \label{W:mass_action_law_working}
\end{equation}
We note that only $\x_\o^{\u\so}$ and $\x_\o^{\u\sobar}$ are the unknown quantities, because, as can be also seen from the above equation, $\x_\o^{\u\s} = 1$ if $\s\notin\{ \so,\sobar \}$, therefore Eq.~(\ref{W:mass_action_law_working}) gives a closed system of $2\n$ equations for $\x_\o^{\u\so}$ and $\x_\o^{\u\sobar}$ at a given pressure. We mention that in the absence of external field, it is also valid that $\x_\o^{\u\sobar}=\x_\obar^{\u\sobar}$, and Eq.~(\ref{W:mass_action_law_working}) corresponds only to $\n$ coupled equations.

To derive the mass action law from the transfer matrix formalism, the dominant eigenvalue from equation Eq.~(\ref{eig.eqs-R}) 
is substituted into equation Eq.~(\ref{eq:x_s^us}), and using the specific form of the transfer matrix, Eq.(\ref{K}),
we have
\begin{align}
     \frac{1}{\x_\o^{\u\so}}
 &=  1 + \frac{\mysum\ov e^{-\beta P\sigma_{\o\ovbar}} \beta P \gamma^{\so\sovbar}
                        \psi^R_\ov}
              {\mysum\ov e^{-\beta P\sigma_{\o\ovbar}} \psi^R_\ov}
  =  1 + \frac{\mysum\ov f_\ov \x_\ov^{\u\sovbar} \beta P \gamma^{\so\sovbar} }
              {\mysum\ov f_\ov \x_\ov^{\u\sovbar}}
\nonumber \\
 &=  1 +\beta P \frac{\mysum\ov f_\ov \x_\ov^{\u\sovbar} \gamma^{\so\sovbar} }
              {\mysum\ov f_\ov \x_\ov^{\u\sov}} ,
  \label{eq:pre_mass_action_law}
\end{align}
where in the second equality we used 
Eq.~(\ref{eq:x_s^usbar/x_sv^uvbar}) and Eq.~(\ref{f_j-trmx}), and in the last equality we used the second equality of Eq.~(\ref{N_mol}).
In a similar way we can derive that
\begin{align}
     \frac{1}{\x_\o^{\u\sobar}}
  =  1 + \beta P \frac{\mysum\ov f_\ov \x_\ov^{\u\sov} {\gamma^{\sov\sobar}} }
              {\mysum\ov f_\ov \x_\ov^{\u\sov}} .
  \label{eq:mass_action_law.2}
\end{align}
We can see that Eqs.~(\ref{eq:pre_mass_action_law}) and (\ref{eq:mass_action_law.2}) are identical with (\ref{W:mass_action_law_final}) for $\s=\so$ and $\s=\sobar$. Finally, when $\s\notin\{ \so,\sobar \}$, it comes from Eq.~(\ref{W:mass_action_law_final}) that  $\x_\o^{\u\s} = 1$, which is trivial in our q1D model and does not need to be proved.

We have shown that both the generalised TPT1 and TMM lead to the same mass action law, Eq.~(\ref{W:mass_action_law_final}) which determines the fraction of sites unbounded at a given pressure. Moreover,  both formalisms lead to Eq.~(\ref{W:f_i}) which determine the orientational distribution function. Finally we show that also the same form of the equation of state can be derived in the framework of the two theories. Since the packing fraction is given by $\eta=\rho\mysum{\o} f_\o d_\o$, therefore the equation of state,  Eq.~(\ref{W:P}),  can be rewritten as
\begin{equation}
     \frac{1}{\rho}
  =  \mysum\o f_\o \biggl(\frac{\x_\o^{\u\so}}{\beta P} + d_\o \biggr) .
  \label{eq:EOS}
\end{equation}
The same equation can be derived from the TMM if we use Eq.~(\ref{eq:<x>}).
Using $f^{(2)}_{\o\ov}(x)$ from Eq.~(\ref{eq:f2(x)_trmx}) we can write that
\begin{align}
    &\int x f^{(2)}_{\o\ov}(x) dx =
 \nonumber \\
 &=  \frac{\psi^L_\o \psi^R_\ov}{\lambda_{\nul}}
     \frac{e^{-\beta(U_\o + P \sigma_{\o\ovbar})}}{\beta P}
     \left[ \frac{1}{\beta P}
           +\sigma_{\o\ovbar} \biggl( 1 + \beta P \gamma^{\so\sovbar}
                        \biggr)
     \right] ,
  \label{pre-EOS_trmx}
\end{align}
therefore, taking into account Eqs.~(\ref{eq:f2_trmx}) and (\ref{eq:f2u_trmx}) we obtain that
\begin{equation}
     \frac{1}{\rho}
  =  \frac{1}{\beta P} \mysum{\o\ov} f^{(2)\uu}_{\o\ov} + \mysum{\o\ov} f^{(2)}_{\o\ov}\sigma_{\o\ovbar}
  =  \mysum\o f_\o \biggl(\frac{\x_\o^{\u\so}}{\beta P} + d_\o \biggr) ,
  \label{EOS_trmx}
\end{equation}
where $\mysum\ov f^{(2)}_{\o\ov} = \mysum\ov f^{(2)}_{\ov\o} = f_\o$ and $\mysum\ov f^{(2)\uu}_{\o\ov} = f_\o^{\uu\so} = f_\o\x_\o^{\u\so}$ are used, moreover, the interaction is additive, $\sigma_{\o\ovbar}=\sigma_\o+\sigma_{\ovbar}$. With the help of Eq.~(\ref{N_mol}) the equation of state can be written in the following simple form
\begin{equation}
     \beta P
  =  \frac{\rho^{\mathrm{mol}}}{1-\eta} .
  \label{EOS_mol}
\end{equation}

Finally, we show that  $e^{\beta\mu}$ can also be expressed by the central quantity of the generalised TPT1 theory, namely by $\x_\o^{\u\s}$. The comparison of Eqs.~(\ref{W:mu(rho,X)}) and (\ref{W:P}) shows that 
\begin{align}
    e^{\beta \mu}
 &= f_\o \frac{\rho}{1-\eta} \x_\o^{\u\so} \x_\o^{\u\sobar} e^{\beta(U_\o + P d_\o)}
\nonumber\\
 &= f_\o \frac{\beta P}{\mysum{\o} f_\o \x_\o^{\u\so}} \x_\o^{\u\so} \x_\o^{\u\sobar} e^{\beta(U_\o + P d_\o)} ,
  \label{e^(beta*mu)}
\end{align}
where $\x_\o^{\u\s} = 1$ if $\s\notin\{ \so,\sobar \}$ is applied. Furthermore, from Eq.~(\ref{W:f_i}) we have 
\begin{equation}
  \frac{f_\o}{\mysum{\ov} f_\ov \x_\ov^{\u\sov}}=\frac{e^{-\beta(U_\o + P d_\o)} /\left(\x_\o^{\u\so} \x_\o^{\u\sobar}\right)}{\mysum{\ov} e^{-\beta(U_\ov + P d_\ov)} / \x_\ov^{\u\sovbar}} .
  \label{help}
\end{equation}
From the combination of Eqs.~(\ref{e^(beta*mu)}) and (\ref{help}), the dominant eigenvalue of the transfer matrix can be expressed as
\begin{equation}
  \lambda_{\nul}=e^{-\beta \mu}=\mysum{\ov} \frac{e^{-\beta(U_\ov + P d_\ov)}}{\beta P \x_\ov^{\u\sovbar}} .
  \label{W:lambda}
\end{equation}
Note that this equation reduces to $\lambda_{\nul}$ of the additive hard body system if all $\x_\o^{\u\s}=1$~\cite{Montero-Santos-Gurin-Varga_JCP_2023}. Moreover, the above result highlights that the key quantities are $P$ and $\x_\o^{\u\s}$.
We note, that Eq~(\ref{W:lambda}) can be obtained also directly from TMM based on Eq.~(\ref{eq:x_s^us}) and then using Eq.~(\ref{eq:x_s^us/x_sv^usv}).

We can also determine the average length of the molecules, $\langle \ell\rangle$, as follows
\begin{equation}
  \mysum{\o} N_{\o} d_{\o}={\cal N}\langle \ell\rangle .
  \label{ell}
\end{equation}
As the number of molecules, ${\cal N}$, is equal to the number of particles with unbounded right site, i.e. ${\cal N}=\mysum{\o} N_{\o}^{\uu\so}$, we get that
\begin{equation}
  \langle\ell\rangle=\frac{\mysum{\o} f_{\o} d_{\o}}{\mysum{\o} f_{\o} \x_{\o}^{\u\so}} .
  \label{ell2}
\end{equation}
Using Eq.~(\ref{W:f_i}) we find the average length as a function of pressure and $\x_\o^{\u\s}$.

In summary, we have managed to prove that generalised TPT1 and TMM are equivalent in thermodynamic limit, where the ensembles are equivalent, so it becomes irrelevant that the generalized TPT1 is defined in $(N\!, L, T)$, while the TMM is defined in  $(N\!, P, T)$ ensembles. In the case of generalised TPT1, instead of solving $3\n$ coupled equations for $f_\o$, $\x_\o^{\u\so}$ and $\x_\o^{\u\sobar}$ $(\o\in\{1,\dots,\n\})$ (Eqs.~(\ref{W:mass_action_law}) and (\ref{W:f})), it is enough to solve only $2\n$ coupled non-linear equations of the mass action law to get $\x_\o^{\u\so}$ and $\x_\o^{\u\sobar}$ (Eq.~(\ref{W:mass_action_law_working})). After that $f_\o$ is given by Eq.~(\ref{W:f_i}), $\rho$ by Eq.~(\ref{eq:EOS}), $\lambda_{\nul}$ by Eq.~(\ref{W:lambda}) and $\langle\ell\rangle$ by Eq.~(\ref{ell2}) at a given pressure. In contrast, in the case of the TMM, the solution of the eigenvalue problem of an $\n\times\n$ matrix provides the dominant eigenvalue, $\lambda_{\nul}$, and the corresponding right and left eigenvectors, $\psi^R_\o$ and  $\psi^L_\o$, respectively. This is a non-trivial, usually numerical task. Having obtained the dominant eigenvalue and the corresponding right and left eigenvectors, the ODF is given by Eq.~(\ref{f_j-trmx}) taking into account the normalisation constraint, $\mysum{\o}f_\o=1$, moreover, we get $\x_\o^{\u\so}$, $\x_\o^{\u\sobar}$ and $\rho$ from Eqs.~(\ref{eq:x_s^us}--\ref{eq:x_s^usbar}) and (\ref{eq:EOS}). These results highlight the fact that the key quantities are the $\x_\o^{\u\so}$, $\x_\o^{\u\sobar}$ and $P$ to study q1D system of sticky hard bodies.

\section{Sticky particles with continuous orientational freedom}

Here we show that the generalization of our results for particles with continuous orientational degrees of freedom is straightforward. In this case the orientation of a particle is described by a two or a three dimensional orientational unit vector, $\ovec$, instead of a discrete orientational index $\o$. We assume an additive contact distance between particles, i.e. $\sigma(\ovec_1,\bar{\ovec}_2) = \sigma(\ovec_1)+\sigma(\bar{\ovec}_2)$, where particle 1 with $\ovec_1$ orientation is on the left side, while particle 2 with $\ovec_2$ on the right side along the channel. Here we follow our previous convention, i.e. $\ovec$ and $\bar{\ovec}$ denote the opposite orientations, when the particle is mirrored and the left and right sides are interchanged, thus $\sigma(\ovec)$ and $\sigma(\bar{\ovec})$ are the right and left ``radius'' of the particle with orientation $\ovec$. We mention that $\sigma(\ovec)$ is not necessarily a shape of a real hard body, it is just an approximate model of a hard-body-like interaction.
Note that among real hard bodies, only spherical shapes exhibit additivity.
Using that $\sigma(\ovec_1,\bar{\ovec}_2)$ is a natural generalization of $\sigma_{\o\ovbar}$, the hard body repulsion can be defined analogously with Eq.~(\ref{eq:V^{HB}}). In a similar way we can define the continuous generalization of the square well attractive interaction energy, $\varepsilon(\svec_1,\svec_2)$. Here we suppose that, as in the discrete case, in each orientation, $\ovec$, the particle has a unique point, denoted by $\svec$, which can be in contact with its right-hand neighbour, irrespectively of the orientation of the neighbouring particle. (Assuming a real hard body shape, this again only occurs in the spherical case.) After this we can write the continuous analogy of Eq.~(\ref{eq:V^{SW}}), then in the sticky limit we can define the stickiness parameter as before in the discrete case, $\gamma(\svec_1,\svec_2) = \lim_{\delta\rightarrow 0}\delta e^{\beta \varepsilon(\svec_1,\svec_2)}$. Finally, using the definition Eq.~(\ref{Delta_def}) we get the analogy of Eq.~(\ref{Delta}):
\begin{align}
\label{Delta_continuous}
     \Delta(\ovec_1,\ovec_2;\svec_3,\svec_4)
  =& \frac{\gamma(\svec_3,\svec_4)}{1-\eta}
         \left[\delta(\svec_3-\svec_1)\delta(\svec_4-\bar{\svec}_2)
         \right.
  \nonumber \\
   &     \left.     
              +\delta(\svec_3-\bar{\svec}_1)\delta(\svec_4-\svec_2)\right] .
\end{align}

Hereafter, all calculations can be repeated analogously to the discrete case, but all sums, $\mysum{\o}$, must be replaced by integrals, $\int d\ovec$. The eigenvalue equation of the transfer matrix become an integral equation, the right and left hand side eigenvectors are functions of $\ovec$, but the results Eqs.~(\ref{f_j-trmx},\ref{eq:f2(x)_trmx}--\ref{eq:<x>})
are the natural generalizations of the previous results.
Similarly, in the generalised TPT1 theory the free energy density, Eq.~(\ref{W:free_energy_density}), becomes a functional of $\rho(\ovec)=\rho f(\ovec)$ and $\x(\ovec_1,\svec_2)$, where $f(\ovec)$ is the orientational distribution function and $\x(\ovec_1,\svec_2)$ is the fraction of sites $\svec_2$ unbounded among the particles with orientation $\ovec_1$.

\section{Conclusion}

We derived the working equations for the orientational ordering and bulk properties of q1D sticky and anisotropic particles, where the particles are confined to a straight line but allowed to rotate and bond in $\n$ different orientational states. With the guidance of the transfer matrix method (TMM), we devised an exact association theory for patchy particles in the sticky limit. The full minimization of the free energy provides the orientational distribution of particles and the fraction of particles not bonded at all $\n$ sites. The free energy contains translational, orientational, packing and association terms separately. Interestingly, the association term is identical to that of Wertheim's first order thermodynamics perturbation theory (TPT1). This means that our association theory can be considered as an generalization of the Wertheim's TPT1 for patchy particles with anisotropic hard core. This is especially true when the orientation of the particles is restricted to a few discrete possibilities, while the additivity of the hard core interaction implies spherical shape in the case of continuous rotational degrees of freedom. What is surprising in these results is that while TPT1 is approximate in higher dimensions with limited applicability, it  can be made exact in one dimension.
There are two crucial conditions in this regard. On the one hand, it is well known that ring formation and double bond formation (when one site bonds to two sites at the same time) limit the applicability of TPT1 theory. These effects are excluded in one dimension, so it is clear, that one of the most important reason of the exactness of the generalized TPT1 in our case is the dimensional restriction. On the other hand, and less intuitively, the stickiness of the interaction is also important for making the theory exact. We have checked that the generalized TPT1 theory is not exact in its present form for a system of particles interacting with a finite range square well potential.

The historical relevance of our work is that we managed to find a link between TMM and TPT1. In the case of TMM, which is devised in isobaric ensemble, it is found that the pressure and the fraction of sites unbounded are the most natural variables, because even the dominant eigenvalue and the corresponding eigenvector of the eigenvalue problem can be expressed analytically with the help of them. In our association theory, which is developed in canonical ensemble, the number of equations for orientational distribution function and the fraction of sites unbounded for $\n$-state sticky particles, can be reduced from $3\n$ to $2\n$ when density is expressed as a function of pressure, the latter being the independent variable.

\begin{figure}
  \centering
  \includegraphics[width=0.95\columnwidth]{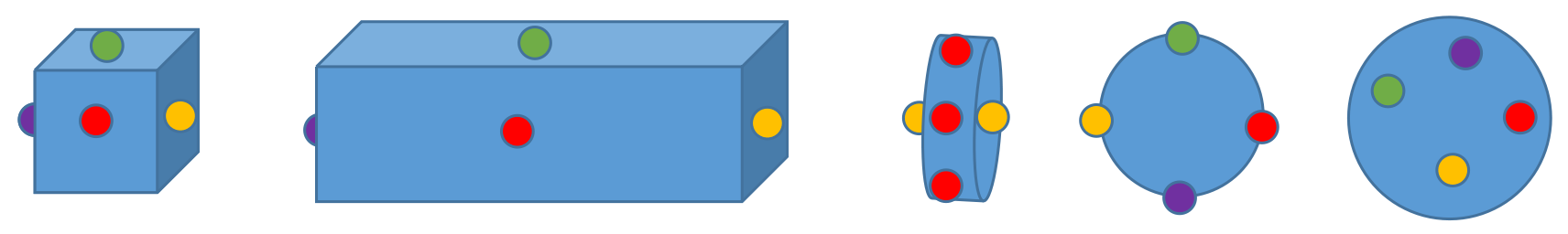}
  \caption{Some orientationally discrete (cube and rectangular board), partially discrete (cylinder) and continuous (2D disk and sphere) hard bodies with some patches whose q1D systems can be studied exactly using TMM or generalised TPT1.}
  \label{fig:general_particle_shapes}
\end{figure}

Our generalised TPT1 together with the TMM opens an avenue for studying exactly the self-assemble and chain formation in q1D associating molecular fluids and patchy colloids. Some possible examples are shown in Fig.~\ref{fig:general_particle_shapes},
where the orientation of the particles can be discrete, partially discrete and continuous. In the discrete case, the maximum number of different sticky systems goes with $2^{\n^2/2+\n/2}$ if patch-patch interactions can be either on or off. This means that the maximum number of different systems is 8 for 2-state, 1024 for 4-state and 2097152 for 6-state systems. We can see from these numbers that it would be a big challenge to study even all possible systems of 4-state sticky particles. Therefore, we must select the interesting systems from the list. It can be shown that dimers, trimers and even $\n$-mers can be formed by a specific choice of bonding interactions. While the dimer formation can be studied with bonds occurring only between the same type of sites (like-like bond), both like-like and unlike-unlike bonds are needed for the formation of $\n$-mers. For example, only monomer, dimers, trimers and 4-mers are allowed to form with one like-like ($\gamma^{22}\neq 0$) and one unlike-unlike ($\gamma^{14}\neq 0$) bonds in the 4-state models (other $\gamma^{\s\sv}=0$). However, the length of the aggregates can grow up to infinity with only one unlike-unlike ($\gamma^{13}\neq 0$) bonds in the same 4 state-model. In the case of $\n$ orientational states, it can be shown that the longest molecule consists of either $\n$ segments or infinite segments.  This shows that the fluid structure of sticky particles can be very rich depending on the type of bonds.

Of course, there are limits of the application of our generalised TPT1 for q1D associating fluids. It is exact only for sticky particles if the bonds are possible along the confining line and the hard body interaction is additive. Although, generalised TPT1 can be extended easily for association sites with finite thickness, it can provide only approximate results. It is also non-trivial to extend it for systems with out-of-line positional fluctuations such as the patchy particles in cylindrical nanotubes. It is also an issue whether the our generalised TPT1 can be extended to be a density functional theory. Based on the work of Kierlik and Rosinberg~\cite{Kierlik-Rosinberg_JStatPhys_1982} and that of Brannock and Percus~\cite{Brannock-Percus_JChemPhys_1996} this is quite feasible as exact functionals are available for two-state sticky hard rods and sticky hard rod mixture with the association term being TPT1-type functional. In this development, the determination of the pair distribution function will be crucial, which can be determined exactly with other 1D methods such as the neighbour distribution method~\cite{Santos_LectureNotesInPhys_2016}.

Regarding the TMM, it is exact in the form presented in this work if only nearest-neighbour interactions are present. However, TMM can be applied for non-additive hard body interactions and finite range bond sites, too. Moreover, bonds can form out-of confining line and one site can bond to two or more other sites in the same time. It can be also extended for out-of-line positional freedoms. Therefore, TMM provides a possible way to get exact results for more realistic associating molecular fluids and patchy colloids in nano-confining environments. We think that using the pressure and the fraction of sites unbonded it will be possible to derive a more general mass action law which is exact for more general cases.

In summary, our work can be a guide for future simulation and density functional theory studies to get reliable results for the association and self-assemble of systems being between one and two dimensions. It is also possible to devise a SAFT-VR type perturbation theory~\cite{10.1063/1.473101}, where the particles are anisotropic and decorated with finite range of association sites and they interact with both short range hard body and attractive interactions.

\section*{Acknowledgements}

Authors gratefully acknowledge the financial support of the National Research, Development, and Innovation Office -- NKFIH K137720 and TKP2021-NKTA-21 and 2023-1.2.4-TÉT-2023-00007. We thank to Patrice Paricaud for the useful discussions.

\bibliography{sticky.bib,/home/gurin/Irodalom/BibTeX/all.bib}

\begin{thebibliography}{76}%
\makeatletter
\providecommand \@ifxundefined [1]{%
 \@ifx{#1\undefined}
}%
\providecommand \@ifnum [1]{%
 \ifnum #1\expandafter \@firstoftwo
 \else \expandafter \@secondoftwo
 \fi
}%
\providecommand \@ifx [1]{%
 \ifx #1\expandafter \@firstoftwo
 \else \expandafter \@secondoftwo
 \fi
}%
\providecommand \natexlab [1]{#1}%
\providecommand \enquote  [1]{``#1''}%
\providecommand \bibnamefont  [1]{#1}%
\providecommand \bibfnamefont [1]{#1}%
\providecommand \citenamefont [1]{#1}%
\providecommand \href@noop [0]{\@secondoftwo}%
\providecommand \href [0]{\begingroup \@sanitize@url \@href}%
\providecommand \@href[1]{\@@startlink{#1}\@@href}%
\providecommand \@@href[1]{\endgroup#1\@@endlink}%
\providecommand \@sanitize@url [0]{\catcode `\\12\catcode `\$12\catcode
  `\&12\catcode `\#12\catcode `\^12\catcode `\_12\catcode `\%12\relax}%
\providecommand \@@startlink[1]{}%
\providecommand \@@endlink[0]{}%
\providecommand \url  [0]{\begingroup\@sanitize@url \@url }%
\providecommand \@url [1]{\endgroup\@href {#1}{\urlprefix }}%
\providecommand \urlprefix  [0]{URL }%
\providecommand \Eprint [0]{\href }%
\providecommand \doibase [0]{https://doi.org/}%
\providecommand \selectlanguage [0]{\@gobble}%
\providecommand \bibinfo  [0]{\@secondoftwo}%
\providecommand \bibfield  [0]{\@secondoftwo}%
\providecommand \translation [1]{[#1]}%
\providecommand \BibitemOpen [0]{}%
\providecommand \bibitemStop [0]{}%
\providecommand \bibitemNoStop [0]{.\EOS\space}%
\providecommand \EOS [0]{\spacefactor3000\relax}%
\providecommand \BibitemShut  [1]{\csname bibitem#1\endcsname}%
\let\auto@bib@innerbib\@empty
\bibitem [{\citenamefont {Glotzer}\ and\ \citenamefont
  {Solomon}(2007)}]{Glotzer-Solomon_NatureMater_2007}%
  \BibitemOpen
  \bibfield  {author} {\bibinfo {author} {\bibfnamefont {S.~C.}\ \bibnamefont
  {Glotzer}}\ and\ \bibinfo {author} {\bibfnamefont {M.~J.}\ \bibnamefont
  {Solomon}},\ }\bibfield  {title} {\bibinfo {title} {Anisotropy of building
  blocks and their assembly into complex structures},\ }\href
  {https://doi.org/10.1038/nmat1949} {\bibfield  {journal} {\bibinfo  {journal}
  {Nat. Mater.}\ }\textbf {\bibinfo {volume} {6}},\ \bibinfo {pages} {557}
  (\bibinfo {year} {2007})}\BibitemShut {NoStop}%
\bibitem [{\citenamefont {Zhang}\ \emph {et~al.}(2014)\citenamefont {Zhang},
  \citenamefont {Regulacio},\ and\ \citenamefont
  {Han}}]{Zhang-Regulacio-Han_ChemSocRev_2014}%
  \BibitemOpen
  \bibfield  {author} {\bibinfo {author} {\bibfnamefont {S.-Y.}\ \bibnamefont
  {Zhang}}, \bibinfo {author} {\bibfnamefont {M.~D.}\ \bibnamefont
  {Regulacio}},\ and\ \bibinfo {author} {\bibfnamefont {M.-Y.}\ \bibnamefont
  {Han}},\ }\bibfield  {title} {\bibinfo {title} {Self-assembly of colloidal
  one-dimensional nanocrystals},\ }\href {https://doi.org/10.1039/C3CS60397K}
  {\bibfield  {journal} {\bibinfo  {journal} {Chem. Soc. Rev.}\ }\textbf
  {\bibinfo {volume} {43}},\ \bibinfo {pages} {2301} (\bibinfo {year}
  {2014})}\BibitemShut {NoStop}%
\bibitem [{\citenamefont {Coluzza}\ \emph {et~al.}(2013)\citenamefont
  {Coluzza}, \citenamefont {van Oostrum}, \citenamefont {Capone}, \citenamefont
  {Reimhult},\ and\ \citenamefont
  {Dellago}}]{Coluzza-vanOostrum-Capone-Reimhult-Dellago_SoftMatter_2013}%
  \BibitemOpen
  \bibfield  {author} {\bibinfo {author} {\bibfnamefont {I.}~\bibnamefont
  {Coluzza}}, \bibinfo {author} {\bibfnamefont {P.~D.~J.}\ \bibnamefont {van
  Oostrum}}, \bibinfo {author} {\bibfnamefont {B.}~\bibnamefont {Capone}},
  \bibinfo {author} {\bibfnamefont {E.}~\bibnamefont {Reimhult}},\ and\
  \bibinfo {author} {\bibfnamefont {C.}~\bibnamefont {Dellago}},\ }\bibfield
  {title} {\bibinfo {title} {Design and folding of colloidal patchy polymers},\
  }\href {https://doi.org/10.1039/C2SM26967H} {\bibfield  {journal} {\bibinfo
  {journal} {Soft Matter}\ }\textbf {\bibinfo {volume} {9}},\ \bibinfo {pages}
  {938} (\bibinfo {year} {2013})}\BibitemShut {NoStop}%
\bibitem [{\citenamefont {Leferink~op Reinink}\ \emph
  {et~al.}(2013)\citenamefont {Leferink~op Reinink}, \citenamefont {van~den
  Pol}, \citenamefont {Petukhov}, \citenamefont {Vroege},\ and\ \citenamefont
  {Lekkerkerker}}]{Leferink.opReinink-van.denPol-Petukhov-Vroege-Lekkerkerker_EurPhysJSpecTop_2013}%
  \BibitemOpen
  \bibfield  {author} {\bibinfo {author} {\bibfnamefont {A.~B. G.~M.}\
  \bibnamefont {Leferink~op Reinink}}, \bibinfo {author} {\bibfnamefont
  {E.}~\bibnamefont {van~den Pol}}, \bibinfo {author} {\bibfnamefont {A.~V.}\
  \bibnamefont {Petukhov}}, \bibinfo {author} {\bibfnamefont {G.~J.}\
  \bibnamefont {Vroege}},\ and\ \bibinfo {author} {\bibfnamefont {H.~N.~W.}\
  \bibnamefont {Lekkerkerker}},\ }\bibfield  {title} {\bibinfo {title} {Phase
  behaviour of lyotropic liquid crystals in external fields and confinement},\
  }\href {https://doi.org/10.1140/epjst/e2013-02075-x} {\bibfield  {journal}
  {\bibinfo  {journal} {Eur. Phys. J. Spec. Top.}\ }\textbf {\bibinfo {volume}
  {222}},\ \bibinfo {pages} {3053–3069} (\bibinfo {year} {2013})}\BibitemShut
  {NoStop}%
\bibitem [{\citenamefont {Royall}\ \emph {et~al.}(2024)\citenamefont {Royall},
  \citenamefont {Charbonneau}, \citenamefont {Dijkstra}, \citenamefont {Russo},
  \citenamefont {Smallenburg}, \citenamefont {Speck},\ and\ \citenamefont
  {Valeriani}}]{Royall-Charbonneau-Dijkstra-Russo-Smallenburg-Speck-Valeriani_RevModPhys_2024}%
  \BibitemOpen
  \bibfield  {author} {\bibinfo {author} {\bibfnamefont {C.~P.}\ \bibnamefont
  {Royall}}, \bibinfo {author} {\bibfnamefont {P.}~\bibnamefont {Charbonneau}},
  \bibinfo {author} {\bibfnamefont {M.}~\bibnamefont {Dijkstra}}, \bibinfo
  {author} {\bibfnamefont {J.}~\bibnamefont {Russo}}, \bibinfo {author}
  {\bibfnamefont {F.}~\bibnamefont {Smallenburg}}, \bibinfo {author}
  {\bibfnamefont {T.}~\bibnamefont {Speck}},\ and\ \bibinfo {author}
  {\bibfnamefont {C.}~\bibnamefont {Valeriani}},\ }\bibfield  {title} {\bibinfo
  {title} {Colloidal hard spheres: Triumphs, challenges, and mysteries},\
  }\href {https://doi.org/10.1103/RevModPhys.96.045003} {\bibfield  {journal}
  {\bibinfo  {journal} {Rev. Mod. Phys.}\ }\textbf {\bibinfo {volume} {96}},\
  \bibinfo {pages} {045003} (\bibinfo {year} {2024})}\BibitemShut {NoStop}%
\bibitem [{\citenamefont {Jack}\ and\ \citenamefont
  {Millett}(2021)}]{Jack-Millett_AIPAdvances_2021}%
  \BibitemOpen
  \bibfield  {author} {\bibinfo {author} {\bibfnamefont {J.~T.}\ \bibnamefont
  {Jack}}\ and\ \bibinfo {author} {\bibfnamefont {P.~C.}\ \bibnamefont
  {Millett}},\ }\bibfield  {title} {\bibinfo {title} {Numerical study of the
  phase behavior of rod-like colloidal particles with attractive tips},\ }\href
  {https://doi.org/10.1063/5.0035565} {\bibfield  {journal} {\bibinfo
  {journal} {AIP Adv.}\ }\textbf {\bibinfo {volume} {11}},\ \bibinfo {pages}
  {025030} (\bibinfo {year} {2021})}\BibitemShut {NoStop}%
\bibitem [{\citenamefont {Kalyuzhnyi}\ \emph {et~al.}(2024)\citenamefont
  {Kalyuzhnyi}, \citenamefont {Patsahan}, \citenamefont {Holovko},\ and\
  \citenamefont
  {Cummings}}]{Kalyuzhnyi-Patsahan-Holovko-Cummings_Nanoscale_2024}%
  \BibitemOpen
  \bibfield  {author} {\bibinfo {author} {\bibfnamefont {Y.~V.}\ \bibnamefont
  {Kalyuzhnyi}}, \bibinfo {author} {\bibfnamefont {T.}~\bibnamefont
  {Patsahan}}, \bibinfo {author} {\bibfnamefont {M.}~\bibnamefont {Holovko}},\
  and\ \bibinfo {author} {\bibfnamefont {P.~T.}\ \bibnamefont {Cummings}},\
  }\bibfield  {title} {\bibinfo {title} {Phase behavior of patchy colloids
  confined in patchy porous media},\ }\href
  {https://doi.org/10.1039/D3NR02866F} {\bibfield  {journal} {\bibinfo
  {journal} {Nanoscale}\ }\textbf {\bibinfo {volume} {16}},\ \bibinfo {pages}
  {4668} (\bibinfo {year} {2024})}\BibitemShut {NoStop}%
\bibitem [{\citenamefont {Vo}\ and\ \citenamefont
  {Glotzer}(2022)}]{Vo-Glotzer_PNAS_2022}%
  \BibitemOpen
  \bibfield  {author} {\bibinfo {author} {\bibfnamefont {T.}~\bibnamefont
  {Vo}}\ and\ \bibinfo {author} {\bibfnamefont {S.~C.}\ \bibnamefont
  {Glotzer}},\ }\bibfield  {title} {\bibinfo {title} {A theory of entropic
  bonding},\ }\href {https://doi.org/10.1073/pnas.2116414119} {\bibfield
  {journal} {\bibinfo  {journal} {PNAS}\ }\textbf {\bibinfo {volume} {119}},\
  \bibinfo {pages} {e2116414119} (\bibinfo {year} {2022})}\BibitemShut
  {NoStop}%
\bibitem [{\citenamefont {Herzfeld}\ and\ \citenamefont
  {Goeppert‐Mayer}(1934)}]{Herzfeld-GoeppertMayer_JChemPhys_1934}%
  \BibitemOpen
  \bibfield  {author} {\bibinfo {author} {\bibfnamefont {K.~F.}\ \bibnamefont
  {Herzfeld}}\ and\ \bibinfo {author} {\bibfnamefont {M.}~\bibnamefont
  {Goeppert‐Mayer}},\ }\bibfield  {title} {\bibinfo {title} {On the states of
  aggregation},\ }\href {https://doi.org/10.1063/1.1749355} {\bibfield
  {journal} {\bibinfo  {journal} {J. Chem. Phys.}\ }\textbf {\bibinfo {volume}
  {2}},\ \bibinfo {pages} {38} (\bibinfo {year} {1934})}\BibitemShut {NoStop}%
\bibitem [{\citenamefont {Brader}\ and\ \citenamefont
  {Evans}(2002)}]{Brader-Evans_PhysicaA_2002}%
  \BibitemOpen
  \bibfield  {author} {\bibinfo {author} {\bibfnamefont {J.}~\bibnamefont
  {Brader}}\ and\ \bibinfo {author} {\bibfnamefont {R.}~\bibnamefont {Evans}},\
  }\bibfield  {title} {\bibinfo {title} {An exactly solvable model for a
  colloid–polymer mixture in one-dimension},\ }\href
  {https://doi.org/https://doi.org/10.1016/S0378-4371(02)00506-X} {\bibfield
  {journal} {\bibinfo  {journal} {Physica A}\ }\textbf {\bibinfo {volume}
  {306}},\ \bibinfo {pages} {287} (\bibinfo {year} {2002})},\ \bibinfo {note}
  {invited Papers from the 21th IUPAP International Conference on Statistical
  Physics}\BibitemShut {NoStop}%
\bibitem [{\citenamefont {Fantoni}\ and\ \citenamefont
  {Santos}(2017)}]{Fantoni-Santos_JStatPhys_2017}%
  \BibitemOpen
  \bibfield  {author} {\bibinfo {author} {\bibfnamefont {R.}~\bibnamefont
  {Fantoni}}\ and\ \bibinfo {author} {\bibfnamefont {A.}~\bibnamefont
  {Santos}},\ }\bibfield  {title} {\bibinfo {title} {One-dimensional fluids
  with second nearest–neighbor interactions},\ }\href
  {https://doi.org/10.1007/s10955-017-1908-6} {\bibfield  {journal} {\bibinfo
  {journal} {J. Stat. Phys.}\ }\textbf {\bibinfo {volume} {169}},\ \bibinfo
  {pages} {1171} (\bibinfo {year} {2017})}\BibitemShut {NoStop}%
\bibitem [{\citenamefont {Montero}\ and\ \citenamefont
  {Santos}(2019)}]{Montero-Santos_JStatPhys_2019}%
  \BibitemOpen
  \bibfield  {author} {\bibinfo {author} {\bibfnamefont {A.~M.}\ \bibnamefont
  {Montero}}\ and\ \bibinfo {author} {\bibfnamefont {A.}~\bibnamefont
  {Santos}},\ }\bibfield  {title} {\bibinfo {title} {Triangle-well and ramp
  interactions in one-dimensional fluids: A fully analytic exact solution},\
  }\href {https://doi.org/10.1007/s10955-019-02255-x} {\bibfield  {journal}
  {\bibinfo  {journal} {J. Stat. Phys.}\ }\textbf {\bibinfo {volume} {175}},\
  \bibinfo {pages} {269} (\bibinfo {year} {2019})}\BibitemShut {NoStop}%
\bibitem [{\citenamefont {Meddour}\ \emph {et~al.}(2024)\citenamefont
  {Meddour}, \citenamefont {Bouzar},\ and\ \citenamefont
  {Messina}}]{Meddour-Bouzar-Messina_JPhysCondensMatter_2025}%
  \BibitemOpen
  \bibfield  {author} {\bibinfo {author} {\bibfnamefont {S.}~\bibnamefont
  {Meddour}}, \bibinfo {author} {\bibfnamefont {L.}~\bibnamefont {Bouzar}},\
  and\ \bibinfo {author} {\bibfnamefont {R.}~\bibnamefont {Messina}},\
  }\bibfield  {title} {\bibinfo {title} {Equation of state and universal solid
  phase of one-dimensional dipolar fluids},\ }\href
  {https://doi.org/10.1088/1361-648X/ad942b} {\bibfield  {journal} {\bibinfo
  {journal} {J. Phys. Condens. Matter}\ }\textbf {\bibinfo {volume} {37}},\
  \bibinfo {pages} {075101} (\bibinfo {year} {2024})}\BibitemShut {NoStop}%
\bibitem [{\citenamefont {Kac}(1959)}]{Kac_PhysFluids_1959}%
  \BibitemOpen
  \bibfield  {author} {\bibinfo {author} {\bibfnamefont {M.}~\bibnamefont
  {Kac}},\ }\bibfield  {title} {\bibinfo {title} {On the partition function of
  a one‐dimensional gas},\ }\href {https://doi.org/10.1063/1.1724399}
  {\bibfield  {journal} {\bibinfo  {journal} {Phys. Fluids}\ }\textbf {\bibinfo
  {volume} {2}},\ \bibinfo {pages} {8} (\bibinfo {year} {1959})}\BibitemShut
  {NoStop}%
\bibitem [{\citenamefont {Carraro}(2003)}]{Carraro_PhysRevE_2003}%
  \BibitemOpen
  \bibfield  {author} {\bibinfo {author} {\bibfnamefont {C.}~\bibnamefont
  {Carraro}},\ }\bibfield  {title} {\bibinfo {title} {Continuous freezing in an
  infinite-range one-dimensional model},\ }\href
  {https://doi.org/10.1103/PhysRevE.67.051502} {\bibfield  {journal} {\bibinfo
  {journal} {Phys. Rev. E}\ }\textbf {\bibinfo {volume} {67}},\ \bibinfo
  {pages} {051502} (\bibinfo {year} {2003})}\BibitemShut {NoStop}%
\bibitem [{\citenamefont {Fantoni}\ \emph {et~al.}(2021)\citenamefont
  {Fantoni}, \citenamefont {Maestre},\ and\ \citenamefont
  {Santos}}]{Fantoni-Maestre-Santos_JStatMech_2021}%
  \BibitemOpen
  \bibfield  {author} {\bibinfo {author} {\bibfnamefont {R.}~\bibnamefont
  {Fantoni}}, \bibinfo {author} {\bibfnamefont {M.~A.~G.}\ \bibnamefont
  {Maestre}},\ and\ \bibinfo {author} {\bibfnamefont {A.}~\bibnamefont
  {Santos}},\ }\bibfield  {title} {\bibinfo {title} {Finite-size effects and
  thermodynamic limit in one-dimensional janus fluids},\ }\href
  {https://doi.org/10.1088/1742-5468/ac2897} {\bibfield  {journal} {\bibinfo
  {journal} {J. Stat. Mech.: Theory Exp.}\ }\textbf {\bibinfo {volume}
  {2021}}\bibinfo  {number} { (10)},\ \bibinfo {pages} {103210}}\BibitemShut
  {NoStop}%
\bibitem [{\citenamefont {Lebowitz}\ \emph {et~al.}(1987)\citenamefont
  {Lebowitz}, \citenamefont {Percus},\ and\ \citenamefont
  {Talbot}}]{Lebowitz-Percus-Talbot_JStatPhys_1987}%
  \BibitemOpen
\bibfield  {number} {  }\bibfield  {author} {\bibinfo {author} {\bibfnamefont
  {J.~L.}\ \bibnamefont {Lebowitz}}, \bibinfo {author} {\bibfnamefont {J.~K.}\
  \bibnamefont {Percus}},\ and\ \bibinfo {author} {\bibfnamefont
  {J.}~\bibnamefont {Talbot}},\ }\bibfield  {title} {\bibinfo {title} {On the
  orientational properties of some one-dimensional model systems},\ }\href
  {https://doi.org/10.1007/BF01017568} {\bibfield  {journal} {\bibinfo
  {journal} {J. Stat. Phys.}\ }\textbf {\bibinfo {volume} {49}},\ \bibinfo
  {pages} {1221} (\bibinfo {year} {1987})}\BibitemShut {NoStop}%
\bibitem [{\citenamefont {Kantor}\ and\ \citenamefont
  {Kardar}(2009)}]{Kantor-Kardar_EPL_2009}%
  \BibitemOpen
  \bibfield  {author} {\bibinfo {author} {\bibfnamefont {Y.}~\bibnamefont
  {Kantor}}\ and\ \bibinfo {author} {\bibfnamefont {M.}~\bibnamefont
  {Kardar}},\ }\bibfield  {title} {\bibinfo {title} {{Universality in the
  jamming limit for elongated hard particles in one dimension}},\ }\href
  {https://doi.org/{10.1209/0295-5075/87/60002}} {\bibfield  {journal}
  {\bibinfo  {journal} {{EPL}}\ }\textbf {\bibinfo {volume} {{87}}},\ \bibinfo
  {pages} {60002} (\bibinfo {year} {{2009}})}\BibitemShut {NoStop}%
\bibitem [{\citenamefont {Joswiak}\ \emph {et~al.}(2016)\citenamefont
  {Joswiak}, \citenamefont {Doherty},\ and\ \citenamefont
  {Peters}}]{Joswiak-Doherty-Peters_JChemPhys_2016}%
  \BibitemOpen
  \bibfield  {author} {\bibinfo {author} {\bibfnamefont {M.~N.}\ \bibnamefont
  {Joswiak}}, \bibinfo {author} {\bibfnamefont {M.~F.}\ \bibnamefont
  {Doherty}},\ and\ \bibinfo {author} {\bibfnamefont {B.}~\bibnamefont
  {Peters}},\ }\bibfield  {title} {\bibinfo {title} {Critical length of a
  one-dimensional nucleus},\ }\href {https://doi.org/10.1063/1.4962448}
  {\bibfield  {journal} {\bibinfo  {journal} {J. Chem. Phys.}\ }\textbf
  {\bibinfo {volume} {145}},\ \bibinfo {pages} {211916} (\bibinfo {year}
  {2016})}\BibitemShut {NoStop}%
\bibitem [{\citenamefont {Bowles}(2000)}]{Bowles_PhysicaA_2000}%
  \BibitemOpen
  \bibfield  {author} {\bibinfo {author} {\bibfnamefont {R.~K.}\ \bibnamefont
  {Bowles}},\ }\bibfield  {title} {\bibinfo {title} {A thermodynamic
  description of the glass transition: an exact one-dimensional example},\
  }\href {https://doi.org/https://doi.org/10.1016/S0378-4371(99)00445-8}
  {\bibfield  {journal} {\bibinfo  {journal} {Physica A}\ }\textbf {\bibinfo
  {volume} {275}},\ \bibinfo {pages} {217} (\bibinfo {year}
  {2000})}\BibitemShut {NoStop}%
\bibitem [{\citenamefont {Godfrey}\ and\ \citenamefont
  {Moore}(2015)}]{Godfrey-Moore_PhysRevE_2015}%
  \BibitemOpen
  \bibfield  {author} {\bibinfo {author} {\bibfnamefont {M.~J.}\ \bibnamefont
  {Godfrey}}\ and\ \bibinfo {author} {\bibfnamefont {M.~A.}\ \bibnamefont
  {Moore}},\ }\bibfield  {title} {\bibinfo {title} {{Understanding the ideal
  glass transition: Lessons from an equilibrium study of hard disks in a
  channel}},\ }\href {https://doi.org/10.1103/PhysRevE.91.022120} {\bibfield
  {journal} {\bibinfo  {journal} {{Physical Review E}}\ }\textbf {\bibinfo
  {volume} {{91}}},\ \bibinfo {pages} {022120} (\bibinfo {year}
  {{2015}})}\BibitemShut {NoStop}%
\bibitem [{\citenamefont {Ashwin}\ and\ \citenamefont
  {Bowles}(2009)}]{Ashwin-Bowles_PRL_2009}%
  \BibitemOpen
  \bibfield  {author} {\bibinfo {author} {\bibfnamefont {S.~S.}\ \bibnamefont
  {Ashwin}}\ and\ \bibinfo {author} {\bibfnamefont {R.~K.}\ \bibnamefont
  {Bowles}},\ }\bibfield  {title} {\bibinfo {title} {Complete jamming landscape
  of confined hard discs},\ }\href
  {https://doi.org/{10.1103/PhysRevLett.102.235701}} {\bibfield  {journal}
  {\bibinfo  {journal} {{Phys. Rev. Lett.}}\ }\textbf {\bibinfo {volume}
  {{102}}},\ \bibinfo {pages} {235701} (\bibinfo {year} {{2009}})}\BibitemShut
  {NoStop}%
\bibitem [{\citenamefont {Zarif}\ \emph {et~al.}(2021)\citenamefont {Zarif},
  \citenamefont {Spiteri},\ and\ \citenamefont
  {Bowles}}]{Zarif-Spiteri-Bowles_PhysRevE_2021}%
  \BibitemOpen
  \bibfield  {author} {\bibinfo {author} {\bibfnamefont {M.}~\bibnamefont
  {Zarif}}, \bibinfo {author} {\bibfnamefont {R.~J.}\ \bibnamefont {Spiteri}},\
  and\ \bibinfo {author} {\bibfnamefont {R.~K.}\ \bibnamefont {Bowles}},\
  }\bibfield  {title} {\bibinfo {title} {Inherent structure landscape of hard
  spheres confined to narrow cylindrical channels},\ }\href
  {https://doi.org/10.1103/PhysRevE.104.064602} {\bibfield  {journal} {\bibinfo
   {journal} {Phys. Rev. E}\ }\textbf {\bibinfo {volume} {104}},\ \bibinfo
  {pages} {064602} (\bibinfo {year} {2021})}\BibitemShut {NoStop}%
\bibitem [{\citenamefont {Salsburg}\ \emph {et~al.}(1953)\citenamefont
  {Salsburg}, \citenamefont {Zwanzig},\ and\ \citenamefont
  {Kirkwood}}]{Salsburg-Zwanzig-Kirkwood_JCP_1953}%
  \BibitemOpen
  \bibfield  {author} {\bibinfo {author} {\bibfnamefont {Z.~W.}\ \bibnamefont
  {Salsburg}}, \bibinfo {author} {\bibfnamefont {R.~W.}\ \bibnamefont
  {Zwanzig}},\ and\ \bibinfo {author} {\bibfnamefont {J.~G.}\ \bibnamefont
  {Kirkwood}},\ }\bibfield  {title} {\bibinfo {title} {Molecular distribution
  functions in a one-dimensional fluid},\ }\href
  {https://doi.org/10.1063/1.1699116} {\bibfield  {journal} {\bibinfo
  {journal} {J. Chem. Phys.}\ }\textbf {\bibinfo {volume} {21}},\ \bibinfo
  {pages} {1098} (\bibinfo {year} {1953})}\BibitemShut {NoStop}%
\bibitem [{\citenamefont {Casey}\ and\ \citenamefont
  {Runnels}(1969)}]{Casey-Runnels_JChemPhys_1969}%
  \BibitemOpen
  \bibfield  {author} {\bibinfo {author} {\bibfnamefont {L.~M.}\ \bibnamefont
  {Casey}}\ and\ \bibinfo {author} {\bibfnamefont {L.~K.}\ \bibnamefont
  {Runnels}},\ }\bibfield  {title} {\bibinfo {title} {Model for correlated
  molecular rotation},\ }\href {https://doi.org/10.1063/1.1671905} {\bibfield
  {journal} {\bibinfo  {journal} {The Journal of Chemical Physics}\ }\textbf
  {\bibinfo {volume} {51}},\ \bibinfo {pages} {5070} (\bibinfo {year}
  {1969})}\BibitemShut {NoStop}%
\bibitem [{\citenamefont {Kofke}\ and\ \citenamefont
  {Post}(1993)}]{Kofke-Post_JCP_1993}%
  \BibitemOpen
  \bibfield  {author} {\bibinfo {author} {\bibfnamefont {D.~A.}\ \bibnamefont
  {Kofke}}\ and\ \bibinfo {author} {\bibfnamefont {A.~J.}\ \bibnamefont
  {Post}},\ }\bibfield  {title} {\bibinfo {title} {{Hard particles in narrow
  pores. Transfer‐matrix solution and the periodic narrow box}},\ }\href
  {https://doi.org/{10.1063/1.464967}} {\bibfield  {journal} {\bibinfo
  {journal} {{J. Chem. Phys.}}\ }\textbf {\bibinfo {volume} {{98}}},\ \bibinfo
  {pages} {4853} (\bibinfo {year} {{1993}})}\BibitemShut {NoStop}%
\bibitem [{\citenamefont {Santos}(2016)}]{Santos_LectureNotesInPhys_2016}%
  \BibitemOpen
  \bibfield  {author} {\bibinfo {author} {\bibfnamefont {A.}~\bibnamefont
  {Santos}},\ }\href {https://doi.org/10.1007/978-3-319-29668-5} {\emph
  {\bibinfo {title} {A Concise Course on the Theory of Classical Liquids.
  Basics and Selected Topics}}},\ \bibinfo {series} {Lecture Notes in Physics},
  Vol.\ \bibinfo {volume} {923}\ (\bibinfo  {publisher} {Springer},\ \bibinfo
  {address} {New York},\ \bibinfo {year} {2016})\BibitemShut {NoStop}%
\bibitem [{\citenamefont {Percus}(2002)}]{Percus_MolPhys_2002}%
  \BibitemOpen
  \bibfield  {author} {\bibinfo {author} {\bibfnamefont {J.~K.}\ \bibnamefont
  {Percus}},\ }\bibfield  {title} {\bibinfo {title} {Density functional theory
  of single-file classical fluids},\ }\href
  {https://doi.org/10.1080/00268970110109925} {\bibfield  {journal} {\bibinfo
  {journal} {Mol. Phys.}\ }\textbf {\bibinfo {volume} {100}},\ \bibinfo {pages}
  {2417} (\bibinfo {year} {2002})}\BibitemShut {NoStop}%
\bibitem [{\citenamefont {Ángel
  Mulero}(2008)}]{Mulero_LectureNotesInPhys_2008}%
  \BibitemOpen
  \bibinfo {editor} {\bibnamefont {Ángel Mulero}},\ ed.,\ \href
  {https://doi.org/10.1007/978-3-540-78767-9} {\emph {\bibinfo {title} {Theory
  and Simulation of Hard-Sphere Fluids and Related Systems}}},\ \bibinfo
  {series} {Lecture Notes in Physics}, Vol.\ \bibinfo {volume} {753}\ (\bibinfo
   {publisher} {Springer},\ \bibinfo {address} {New York},\ \bibinfo {year}
  {2008})\BibitemShut {NoStop}%
\bibitem [{\citenamefont {Mederos}\ \emph {et~al.}(2014)\citenamefont
  {Mederos}, \citenamefont {Velasco},\ and\ \citenamefont
  {Mart\'{\i}nez-Rat\'on}}]{Luis-Enrique-Yuri_JPhysCondMat_2014}%
  \BibitemOpen
  \bibfield  {author} {\bibinfo {author} {\bibfnamefont {L.}~\bibnamefont
  {Mederos}}, \bibinfo {author} {\bibfnamefont {E.}~\bibnamefont {Velasco}},\
  and\ \bibinfo {author} {\bibfnamefont {Y.}~\bibnamefont
  {Mart\'{\i}nez-Rat\'on}},\ }\bibfield  {title} {\bibinfo {title} {Hard-body
  models of bulk liquid crystals},\ }\href
  {https://doi.org/10.1088/0953-8984/26/46/463101} {\bibfield  {journal}
  {\bibinfo  {journal} {J. Phys.: Condens. Matter}\ }\textbf {\bibinfo {volume}
  {26}},\ \bibinfo {pages} {463101} (\bibinfo {year} {2014})}\BibitemShut
  {NoStop}%
\bibitem [{\citenamefont {Percus}(1976)}]{Percus_JStatPhys_1976}%
  \BibitemOpen
  \bibfield  {author} {\bibinfo {author} {\bibfnamefont {J.~K.}\ \bibnamefont
  {Percus}},\ }\bibfield  {title} {\bibinfo {title} {Equilibrium state of a
  classical fluid of hard rods in an external field},\ }\href
  {https://doi.org/10.1007/BF01020803} {\bibfield  {journal} {\bibinfo
  {journal} {J. Stat. Phys.}\ }\textbf {\bibinfo {volume} {15}},\ \bibinfo
  {pages} {505} (\bibinfo {year} {1976})}\BibitemShut {NoStop}%
\bibitem [{\citenamefont {Percus}(1982)}]{Percus_JStatPhys_1982}%
  \BibitemOpen
  \bibfield  {author} {\bibinfo {author} {\bibfnamefont {J.~K.}\ \bibnamefont
  {Percus}},\ }\bibfield  {title} {\bibinfo {title} {One-dimensional classical
  fluid with nearest-neighbor interaction in arbitrary external field},\ }\href
  {https://doi.org/10.1007/BF01011623} {\bibfield  {journal} {\bibinfo
  {journal} {J. Stat. Phys.}\ }\textbf {\bibinfo {volume} {28}},\ \bibinfo
  {pages} {67} (\bibinfo {year} {1982})}\BibitemShut {NoStop}%
\bibitem [{\citenamefont {Kierlik}\ and\ \citenamefont
  {Rosinberg}(1992)}]{Kierlik-Rosinberg_JStatPhys_1982}%
  \BibitemOpen
  \bibfield  {author} {\bibinfo {author} {\bibfnamefont {E.}~\bibnamefont
  {Kierlik}}\ and\ \bibinfo {author} {\bibfnamefont {M.~L.}\ \bibnamefont
  {Rosinberg}},\ }\bibfield  {title} {\bibinfo {title} {The classical fluid of
  associating hard rods in an external field},\ }\href
  {https://doi.org/10.1007/BF01048884} {\bibfield  {journal} {\bibinfo
  {journal} {J. Stat. Phys.}\ }\textbf {\bibinfo {volume} {68}},\ \bibinfo
  {pages} {1037} (\bibinfo {year} {1992})}\BibitemShut {NoStop}%
\bibitem [{\citenamefont {Vanderlick}\ \emph {et~al.}(1989)\citenamefont
  {Vanderlick}, \citenamefont {Davis},\ and\ \citenamefont
  {Percus}}]{Vanderlick-Davis-Percus_JChemPhys_1989}%
  \BibitemOpen
  \bibfield  {author} {\bibinfo {author} {\bibfnamefont {T.~K.}\ \bibnamefont
  {Vanderlick}}, \bibinfo {author} {\bibfnamefont {H.~T.}\ \bibnamefont
  {Davis}},\ and\ \bibinfo {author} {\bibfnamefont {J.~K.}\ \bibnamefont
  {Percus}},\ }\bibfield  {title} {\bibinfo {title} {The statistical mechanics
  of inhomogeneous hard rod mixtures},\ }\href
  {https://doi.org/10.1063/1.457329} {\bibfield  {journal} {\bibinfo  {journal}
  {J. Chem. Phys.}\ }\textbf {\bibinfo {volume} {91}},\ \bibinfo {pages} {7136}
  (\bibinfo {year} {1989})}\BibitemShut {NoStop}%
\bibitem [{\citenamefont {Brannock}\ and\ \citenamefont
  {Percus}(1996)}]{Brannock-Percus_JChemPhys_1996}%
  \BibitemOpen
  \bibfield  {author} {\bibinfo {author} {\bibfnamefont {G.~R.}\ \bibnamefont
  {Brannock}}\ and\ \bibinfo {author} {\bibfnamefont {J.~K.}\ \bibnamefont
  {Percus}},\ }\bibfield  {title} {\bibinfo {title} {Wertheim cluster
  development of free energy functionals for general nearest‐neighbor
  interactions in d=1},\ }\href {https://doi.org/10.1063/1.471920} {\bibfield
  {journal} {\bibinfo  {journal} {J. Chem. Phys.}\ }\textbf {\bibinfo {volume}
  {105}},\ \bibinfo {pages} {614} (\bibinfo {year} {1996})}\BibitemShut
  {NoStop}%
\bibitem [{\citenamefont {Percus}(1997)}]{Percus_JStatPhys_1997}%
  \BibitemOpen
  \bibfield  {author} {\bibinfo {author} {\bibfnamefont {J.~K.}\ \bibnamefont
  {Percus}},\ }\bibfield  {title} {\bibinfo {title} {Nonuniform classical fluid
  mixture in one-dimensional space with next neighbor interactions},\ }\href
  {https://doi.org/10.1007/BF02770764} {\bibfield  {journal} {\bibinfo
  {journal} {J. Stat. Phys.}\ }\textbf {\bibinfo {volume} {89}},\ \bibinfo
  {pages} {249} (\bibinfo {year} {1997})}\BibitemShut {NoStop}%
\bibitem [{\citenamefont {Tutschka}\ and\ \citenamefont
  {Cuesta}(2003)}]{Tutschka-Cuesta_JStatPhys_2003}%
  \BibitemOpen
  \bibfield  {author} {\bibinfo {author} {\bibfnamefont {C.}~\bibnamefont
  {Tutschka}}\ and\ \bibinfo {author} {\bibfnamefont {J.~A.}\ \bibnamefont
  {Cuesta}},\ }\bibfield  {title} {\bibinfo {title} {Overcomplete free energy
  functional for {$D=1$} particle systems with next neighbor interactions},\
  }\href {https://doi.org/10.1023/A:1023096031180} {\bibfield  {journal}
  {\bibinfo  {journal} {J. Stat. Phys.}\ }\textbf {\bibinfo {volume} {111}},\
  \bibinfo {pages} {1125} (\bibinfo {year} {2003})}\BibitemShut {NoStop}%
\bibitem [{\citenamefont {Sahnoun}\ \emph {et~al.}(2024)\citenamefont
  {Sahnoun}, \citenamefont {Djebbar}, \citenamefont {Benmessabih},\ and\
  \citenamefont
  {Bakhti}}]{Sahnoun-Djebbar-Benmessabih-Bakhti_JPhysA-MathTheor_2024}%
  \BibitemOpen
  \bibfield  {author} {\bibinfo {author} {\bibfnamefont {A.~Y.}\ \bibnamefont
  {Sahnoun}}, \bibinfo {author} {\bibfnamefont {M.}~\bibnamefont {Djebbar}},
  \bibinfo {author} {\bibfnamefont {T.}~\bibnamefont {Benmessabih}},\ and\
  \bibinfo {author} {\bibfnamefont {B.}~\bibnamefont {Bakhti}},\ }\bibfield
  {title} {\bibinfo {title} {One dimensional lattice fluid mixture with nearest
  neighbour interactions},\ }\href {https://doi.org/10.1088/1751-8121/ad6538}
  {\bibfield  {journal} {\bibinfo  {journal} {J. Phys. A -- Math. Theor.}\
  }\textbf {\bibinfo {volume} {57}},\ \bibinfo {pages} {325007} (\bibinfo
  {year} {2024})}\BibitemShut {NoStop}%
\bibitem [{\citenamefont {Rosenfeld}(1989)}]{Rosenfeld_PRL_1989}%
  \BibitemOpen
  \bibfield  {author} {\bibinfo {author} {\bibfnamefont {Y.}~\bibnamefont
  {Rosenfeld}},\ }\bibfield  {title} {\bibinfo {title} {Free-energy model for
  the inhomogeneous hard-sphere fluid mixture and density-functional theory of
  freezing},\ }\href {https://doi.org/10.1103/PhysRevLett.63.980} {\bibfield
  {journal} {\bibinfo  {journal} {Phys. Rev. Lett.}\ }\textbf {\bibinfo
  {volume} {63}},\ \bibinfo {pages} {980} (\bibinfo {year} {1989})}\BibitemShut
  {NoStop}%
\bibitem [{\citenamefont {Lin}\ and\ \citenamefont
  {Oettel}(2019)}]{Lin-Oettel_SciPostPhys_2019}%
  \BibitemOpen
  \bibfield  {author} {\bibinfo {author} {\bibfnamefont {S.-C.}\ \bibnamefont
  {Lin}}\ and\ \bibinfo {author} {\bibfnamefont {M.}~\bibnamefont {Oettel}},\
  }\bibfield  {title} {\bibinfo {title} {{A classical density functional from
  machine learning and a convolutional neural network}},\ }\href
  {https://doi.org/10.21468/SciPostPhys.6.2.025} {\bibfield  {journal}
  {\bibinfo  {journal} {SciPost Phys.}\ }\textbf {\bibinfo {volume} {6}},\
  \bibinfo {pages} {025} (\bibinfo {year} {2019})}\BibitemShut {NoStop}%
\bibitem [{\citenamefont {Wu}\ and\ \citenamefont
  {Li}(2007)}]{Wu-Li_AnnuRevPhysChem_2007}%
  \BibitemOpen
  \bibfield  {author} {\bibinfo {author} {\bibfnamefont {J.}~\bibnamefont
  {Wu}}\ and\ \bibinfo {author} {\bibfnamefont {Z.}~\bibnamefont {Li}},\
  }\bibfield  {title} {\bibinfo {title} {Density-functional theory for complex
  fluids},\ }\href
  {https://doi.org/https://doi.org/10.1146/annurev.physchem.58.032806.104650}
  {\bibfield  {journal} {\bibinfo  {journal} {Annu. Rev. Phys. Chem.}\ }\textbf
  {\bibinfo {volume} {58}},\ \bibinfo {pages} {85} (\bibinfo {year}
  {2007})}\BibitemShut {NoStop}%
\bibitem [{\citenamefont {Schmidt}(2007)}]{Schmidt_PRE_2007}%
  \BibitemOpen
  \bibfield  {author} {\bibinfo {author} {\bibfnamefont {M.}~\bibnamefont
  {Schmidt}},\ }\bibfield  {title} {\bibinfo {title} {Fundamental measure
  density functional theory for nonadditive hard-core mixtures: The
  one-dimensional case},\ }\href {https://doi.org/10.1103/PhysRevE.76.031202}
  {\bibfield  {journal} {\bibinfo  {journal} {Phys. Rev. E}\ }\textbf {\bibinfo
  {volume} {76}},\ \bibinfo {pages} {031202} (\bibinfo {year}
  {2007})}\BibitemShut {NoStop}%
\bibitem [{\citenamefont {El~Moumane}\ \emph {et~al.}(2024)\citenamefont
  {El~Moumane}, \citenamefont {te~Vrugt}, \citenamefont {Löwen},\ and\
  \citenamefont {Wittmann}}]{ElMoumane-teVrugt-Loewen-Wittmann_JChemPhys_2024}%
  \BibitemOpen
  \bibfield  {author} {\bibinfo {author} {\bibfnamefont {A.}~\bibnamefont
  {El~Moumane}}, \bibinfo {author} {\bibfnamefont {M.}~\bibnamefont
  {te~Vrugt}}, \bibinfo {author} {\bibfnamefont {H.}~\bibnamefont {Löwen}},\
  and\ \bibinfo {author} {\bibfnamefont {R.}~\bibnamefont {Wittmann}},\
  }\bibfield  {title} {\bibinfo {title} {Biaxial nematic order in fundamental
  measure theory},\ }\href {https://doi.org/10.1063/5.0188117} {\bibfield
  {journal} {\bibinfo  {journal} {J. Chem. Phys.}\ }\textbf {\bibinfo {volume}
  {160}},\ \bibinfo {pages} {094903} (\bibinfo {year} {2024})}\BibitemShut
  {NoStop}%
\bibitem [{\citenamefont {Marshall}(2015)}]{Marshall_JChemPhys_2015}%
  \BibitemOpen
  \bibfield  {author} {\bibinfo {author} {\bibfnamefont {B.~D.}\ \bibnamefont
  {Marshall}},\ }\bibfield  {title} {\bibinfo {title} {Thermodynamic
  perturbation theory for associating fluids confined in a one-dimensional
  pore},\ }\href {https://doi.org/10.1063/1.4922547} {\bibfield  {journal}
  {\bibinfo  {journal} {J. Chem. Phys.}\ }\textbf {\bibinfo {volume} {142}},\
  \bibinfo {pages} {234906} (\bibinfo {year} {2015})}\BibitemShut {NoStop}%
\bibitem [{\citenamefont {Marshall}(2016)}]{Marshall_PhysRevE_2016}%
  \BibitemOpen
  \bibfield  {author} {\bibinfo {author} {\bibfnamefont {B.~D.}\ \bibnamefont
  {Marshall}},\ }\bibfield  {title} {\bibinfo {title} {Equilibrium adsorption
  and self-assembly of patchy colloids in microchannels},\ }\href
  {https://doi.org/10.1103/PhysRevE.94.012615} {\bibfield  {journal} {\bibinfo
  {journal} {Phys. Rev. E}\ }\textbf {\bibinfo {volume} {94}},\ \bibinfo
  {pages} {012615} (\bibinfo {year} {2016})}\BibitemShut {NoStop}%
\bibitem [{\citenamefont {Percus}\ and\ \citenamefont
  {Zhang}(1990)}]{Percus-Zhang_MolPhys_1990}%
  \BibitemOpen
  \bibfield  {author} {\bibinfo {author} {\bibfnamefont {J.}~\bibnamefont
  {Percus}}\ and\ \bibinfo {author} {\bibfnamefont {M.}~\bibnamefont {Zhang}},\
  }\bibfield  {title} {\bibinfo {title} {The quasi-one dimensional hard square
  gas},\ }\href {https://doi.org/10.1080/00268979000100241} {\bibfield
  {journal} {\bibinfo  {journal} {Mol. Phys.}\ }\textbf {\bibinfo {volume}
  {69}},\ \bibinfo {pages} {347} (\bibinfo {year} {1990})}\BibitemShut
  {NoStop}%
\bibitem [{\citenamefont {Gurin}\ and\ \citenamefont
  {Varga}(2015)}]{Gurin-Varga_JCP_2015}%
  \BibitemOpen
  \bibfield  {author} {\bibinfo {author} {\bibfnamefont {P.}~\bibnamefont
  {Gurin}}\ and\ \bibinfo {author} {\bibfnamefont {S.}~\bibnamefont {Varga}},\
  }\bibfield  {title} {\bibinfo {title} {Beyond the single-file fluid limit
  using transfer matrix method: Exact results for confined parallel hard
  squares},\ }\href {https://doi.org/10.1063/1.4922154} {\bibfield  {journal}
  {\bibinfo  {journal} {J. Chem. Phys.}\ }\textbf {\bibinfo {volume} {142}},\
  \bibinfo {pages} {224503} (\bibinfo {year} {2015})}\BibitemShut {NoStop}%
\bibitem [{\citenamefont {Lin}\ \emph {et~al.}(2009)\citenamefont {Lin},
  \citenamefont {Valley}, \citenamefont {Meron}, \citenamefont {Cui},
  \citenamefont {Ho},\ and\ \citenamefont
  {Rice}}]{Lin-Valley-Meron-Cui-Ho-Rice_JPhysChemB_2009}%
  \BibitemOpen
  \bibfield  {author} {\bibinfo {author} {\bibfnamefont {B.}~\bibnamefont
  {Lin}}, \bibinfo {author} {\bibfnamefont {D.}~\bibnamefont {Valley}},
  \bibinfo {author} {\bibfnamefont {M.}~\bibnamefont {Meron}}, \bibinfo
  {author} {\bibfnamefont {B.}~\bibnamefont {Cui}}, \bibinfo {author}
  {\bibfnamefont {H.~M.}\ \bibnamefont {Ho}},\ and\ \bibinfo {author}
  {\bibfnamefont {S.~A.}\ \bibnamefont {Rice}},\ }\bibfield  {title} {\bibinfo
  {title} {The quasi-one-dimensional colloid fluid revisited},\ }\href
  {https://doi.org/10.1021/jp9018734} {\bibfield  {journal} {\bibinfo
  {journal} {J. Phys. Chem. B}\ }\textbf {\bibinfo {volume} {113}},\ \bibinfo
  {pages} {13742} (\bibinfo {year} {2009})},\ \bibinfo {note} {pMID:
  19569626}\BibitemShut {NoStop}%
\bibitem [{\citenamefont {Gorantla}\ \emph {et~al.}(2010)\citenamefont
  {Gorantla}, \citenamefont {Börrnert}, \citenamefont {Bachmatiuk},
  \citenamefont {Dimitrakopoulou}, \citenamefont {Schönfelder}, \citenamefont
  {Schäffel}, \citenamefont {Thomas}, \citenamefont {Gemming}, \citenamefont
  {Borowiak-Palen}, \citenamefont {Warner}, \citenamefont {Yakobson},
  \citenamefont {Eckert}, \citenamefont {Büchner},\ and\ \citenamefont
  {Rümmeli}}]{Gorantla_et.al_Nanoscale_2010}%
  \BibitemOpen
  \bibfield  {author} {\bibinfo {author} {\bibfnamefont {S.}~\bibnamefont
  {Gorantla}}, \bibinfo {author} {\bibfnamefont {F.}~\bibnamefont {Börrnert}},
  \bibinfo {author} {\bibfnamefont {A.}~\bibnamefont {Bachmatiuk}}, \bibinfo
  {author} {\bibfnamefont {M.}~\bibnamefont {Dimitrakopoulou}}, \bibinfo
  {author} {\bibfnamefont {R.}~\bibnamefont {Schönfelder}}, \bibinfo {author}
  {\bibfnamefont {F.}~\bibnamefont {Schäffel}}, \bibinfo {author}
  {\bibfnamefont {J.}~\bibnamefont {Thomas}}, \bibinfo {author} {\bibfnamefont
  {T.}~\bibnamefont {Gemming}}, \bibinfo {author} {\bibfnamefont
  {E.}~\bibnamefont {Borowiak-Palen}}, \bibinfo {author} {\bibfnamefont
  {J.~H.}\ \bibnamefont {Warner}}, \bibinfo {author} {\bibfnamefont {B.~I.}\
  \bibnamefont {Yakobson}}, \bibinfo {author} {\bibfnamefont {J.}~\bibnamefont
  {Eckert}}, \bibinfo {author} {\bibfnamefont {B.}~\bibnamefont {Büchner}},\
  and\ \bibinfo {author} {\bibfnamefont {M.~H.}\ \bibnamefont {Rümmeli}},\
  }\bibfield  {title} {\bibinfo {title} {In situ observations of fullerene
  fusion and ejection in carbon nanotubes},\ }\href
  {https://doi.org/10.1039/C0NR00426J} {\bibfield  {journal} {\bibinfo
  {journal} {Nanoscale}\ }\textbf {\bibinfo {volume} {2}},\ \bibinfo {pages}
  {2077} (\bibinfo {year} {2010})}\BibitemShut {NoStop}%
\bibitem [{\citenamefont {Ashwin}\ \emph {et~al.}(2013)\citenamefont {Ashwin},
  \citenamefont {Yamchi},\ and\ \citenamefont
  {Bowles}}]{Ashwin-Yamchi-Bowles_PhysRevLett_2013}%
  \BibitemOpen
  \bibfield  {author} {\bibinfo {author} {\bibfnamefont {S.~S.}\ \bibnamefont
  {Ashwin}}, \bibinfo {author} {\bibfnamefont {M.~Z.}\ \bibnamefont {Yamchi}},\
  and\ \bibinfo {author} {\bibfnamefont {R.~K.}\ \bibnamefont {Bowles}},\
  }\bibfield  {title} {\bibinfo {title} {Inherent structure landscape
  connection between liquids, granular materials, and the jamming phase
  diagram},\ }\href {https://doi.org/{10.1103/PhysRevLett.110.145701}}
  {\bibfield  {journal} {\bibinfo  {journal} {{Phys. Rev. Lett.}}\ }\textbf
  {\bibinfo {volume} {{110}}},\ \bibinfo {pages} {145701} (\bibinfo {year}
  {{2013}})}\BibitemShut {NoStop}%
\bibitem [{\citenamefont {Robinson}\ \emph {et~al.}(2016)\citenamefont
  {Robinson}, \citenamefont {Godfrey},\ and\ \citenamefont
  {Moore}}]{Robinson-Godfrey-Moore_PhysRevE_2016}%
  \BibitemOpen
  \bibfield  {author} {\bibinfo {author} {\bibfnamefont {J.~F.}\ \bibnamefont
  {Robinson}}, \bibinfo {author} {\bibfnamefont {M.~J.}\ \bibnamefont
  {Godfrey}},\ and\ \bibinfo {author} {\bibfnamefont {M.~A.}\ \bibnamefont
  {Moore}},\ }\bibfield  {title} {\bibinfo {title} {{Glasslike behavior of a
  hard-disk fluid confined to a narrow channel}},\ }\href
  {https://doi.org/10.1103/PhysRevE.93.032101} {\bibfield  {journal} {\bibinfo
  {journal} {{Physical Review E}}\ }\textbf {\bibinfo {volume} {{93}}},\
  \bibinfo {pages} {032101} (\bibinfo {year} {{2016}})}\BibitemShut {NoStop}%
\bibitem [{\citenamefont {Montero}\ \emph {et~al.}(2023)\citenamefont
  {Montero}, \citenamefont {Santos}, \citenamefont {Gurin},\ and\ \citenamefont
  {Varga}}]{Montero-Santos-Gurin-Varga_JCP_2023}%
  \BibitemOpen
  \bibfield  {author} {\bibinfo {author} {\bibfnamefont {A.~M.}\ \bibnamefont
  {Montero}}, \bibinfo {author} {\bibfnamefont {A.}~\bibnamefont {Santos}},
  \bibinfo {author} {\bibfnamefont {P.}~\bibnamefont {Gurin}},\ and\ \bibinfo
  {author} {\bibfnamefont {S.}~\bibnamefont {Varga}},\ }\bibfield  {title}
  {\bibinfo {title} {{Ordering properties of anisotropic hard bodies in
  one-dimensional channels}},\ }\href {https://doi.org/10.1063/5.0169605}
  {\bibfield  {journal} {\bibinfo  {journal} {The Journal of Chemical Physics}\
  }\textbf {\bibinfo {volume} {159}},\ \bibinfo {pages} {154507} (\bibinfo
  {year} {2023})}\BibitemShut {NoStop}%
\bibitem [{\citenamefont {Gurin}\ \emph {et~al.}(2024)\citenamefont {Gurin},
  \citenamefont {Mizani},\ and\ \citenamefont
  {Varga}}]{Gurin-Mizani-Varga_PhysRevE_2024}%
  \BibitemOpen
  \bibfield  {author} {\bibinfo {author} {\bibfnamefont {P.}~\bibnamefont
  {Gurin}}, \bibinfo {author} {\bibfnamefont {S.}~\bibnamefont {Mizani}},\ and\
  \bibinfo {author} {\bibfnamefont {S.}~\bibnamefont {Varga}},\ }\bibfield
  {title} {\bibinfo {title} {Orientational ordering and correlation in
  quasi-one-dimensional hard-body fluids due to close-packing degeneracy},\
  }\href {https://doi.org/10.1103/PhysRevE.110.014702} {\bibfield  {journal}
  {\bibinfo  {journal} {Phys. Rev. E}\ }\textbf {\bibinfo {volume} {110}},\
  \bibinfo {pages} {014702} (\bibinfo {year} {2024})}\BibitemShut {NoStop}%
\bibitem [{\citenamefont {Mizani}\ \emph
  {et~al.}(2025{\natexlab{a}})\citenamefont {Mizani}, \citenamefont {Oettel},
  \citenamefont {Gurin},\ and\ \citenamefont
  {Varga}}]{Mizani-Oettel-Gurin-Varga_SciPostPhysCore_2025}%
  \BibitemOpen
  \bibfield  {author} {\bibinfo {author} {\bibfnamefont {S.}~\bibnamefont
  {Mizani}}, \bibinfo {author} {\bibfnamefont {M.}~\bibnamefont {Oettel}},
  \bibinfo {author} {\bibfnamefont {P.}~\bibnamefont {Gurin}},\ and\ \bibinfo
  {author} {\bibfnamefont {S.}~\bibnamefont {Varga}},\ }\bibfield  {title}
  {\bibinfo {title} {Universality of the close packing properties and markers
  of isotropic-to-tetratic crossover in quasi-one-dimensional superdisk
  fluid},\ }\href {https://doi.org/10.21468/SciPostPhysCore.8.1.008} {\bibfield
   {journal} {\bibinfo  {journal} {SciPost Phys. Core}\ }\textbf {\bibinfo
  {volume} {8}},\ \bibinfo {pages} {008} (\bibinfo {year}
  {2025}{\natexlab{a}})}\BibitemShut {NoStop}%
\bibitem [{\citenamefont {Mizani}\ \emph
  {et~al.}(2025{\natexlab{b}})\citenamefont {Mizani}, \citenamefont {Gurin},
  \citenamefont {Varga},\ and\ \citenamefont
  {Oettel}}]{Mizani-Gurin-Varga-Oettel_PhysRevE_2025}%
  \BibitemOpen
  \bibfield  {author} {\bibinfo {author} {\bibfnamefont {S.}~\bibnamefont
  {Mizani}}, \bibinfo {author} {\bibfnamefont {P.}~\bibnamefont {Gurin}},
  \bibinfo {author} {\bibfnamefont {S.}~\bibnamefont {Varga}},\ and\ \bibinfo
  {author} {\bibfnamefont {M.}~\bibnamefont {Oettel}},\ }\bibfield  {title}
  {\bibinfo {title} {Competition between shape anisotropy and deformation in
  the ordering and close packing properties of quasi-one-dimensional hard
  superellipse fluids},\ }\bibfield  {journal} {\bibinfo  {journal} {accepted
  in Phys. Rev. E}\ }\href {https://doi.org/doi.org/10.1103/jszf-cvdn}
  {doi.org/10.1103/jszf-cvdn} (\bibinfo {year}
  {2025}{\natexlab{b}})\BibitemShut {NoStop}%
\bibitem [{\citenamefont {Lebowitz}\ and\ \citenamefont
  {Percus}(1983)}]{Lebowitz-Percus_AnnNYAcadSci_1983}%
  \BibitemOpen
  \bibfield  {author} {\bibinfo {author} {\bibfnamefont {J.~L.}\ \bibnamefont
  {Lebowitz}}\ and\ \bibinfo {author} {\bibfnamefont {J.~K.}\ \bibnamefont
  {Percus}},\ }\bibfield  {title} {\bibinfo {title} {One-dimensional models of
  anisotropic fluids},\ }\href
  {https://doi.org/10.1111/j.1749-6632.1983.tb23333.x} {\bibfield  {journal}
  {\bibinfo  {journal} {Ann. N. Y. Acad. Sci.}\ }\textbf {\bibinfo {volume}
  {410}},\ \bibinfo {pages} {351} (\bibinfo {year} {1983})}\BibitemShut
  {NoStop}%
\bibitem [{\citenamefont {Gurin}\ and\ \citenamefont
  {Varga}(2022)}]{Gurin-Varga_PhysRevE_2022}%
  \BibitemOpen
  \bibfield  {author} {\bibinfo {author} {\bibfnamefont {P.}~\bibnamefont
  {Gurin}}\ and\ \bibinfo {author} {\bibfnamefont {S.}~\bibnamefont {Varga}},\
  }\bibfield  {title} {\bibinfo {title} {Anomalous phase behavior of
  quasi-one-dimensional attractive hard rods},\ }\href
  {https://doi.org/10.1103/PhysRevE.106.044606} {\bibfield  {journal} {\bibinfo
   {journal} {Phys. Rev. E}\ }\textbf {\bibinfo {volume} {106}},\ \bibinfo
  {pages} {044606} (\bibinfo {year} {2022})}\BibitemShut {NoStop}%
\bibitem [{\citenamefont {Baxter}(1968)}]{Baxter_JChemPhys_1968}%
  \BibitemOpen
  \bibfield  {author} {\bibinfo {author} {\bibfnamefont {R.~J.}\ \bibnamefont
  {Baxter}},\ }\bibfield  {title} {\bibinfo {title} {Percus–yevick equation
  for hard spheres with surface adhesion},\ }\href
  {https://doi.org/10.1063/1.1670482} {\bibfield  {journal} {\bibinfo
  {journal} {J. Chem. Phys.}\ }\textbf {\bibinfo {volume} {49}},\ \bibinfo
  {pages} {2770} (\bibinfo {year} {1968})}\BibitemShut {NoStop}%
\bibitem [{\citenamefont
  {Wertheim}(1984{\natexlab{a}})}]{Wertheim_JStatPhys_1984}%
  \BibitemOpen
  \bibfield  {author} {\bibinfo {author} {\bibfnamefont {M.~S.}\ \bibnamefont
  {Wertheim}},\ }\bibfield  {title} {\bibinfo {title} {Fluids with highly
  directional attractive forces. {I}. statistical thermodynamics},\ }\href
  {https://doi.org/10.1007/BF01017362} {\bibfield  {journal} {\bibinfo
  {journal} {J. Stat. Phys.}\ }\textbf {\bibinfo {volume} {35}},\ \bibinfo
  {pages} {19} (\bibinfo {year} {1984}{\natexlab{a}})}\BibitemShut {NoStop}%
\bibitem [{\citenamefont
  {Wertheim}(1984{\natexlab{b}})}]{Wertheim.1_JStatPhys_1984}%
  \BibitemOpen
  \bibfield  {author} {\bibinfo {author} {\bibfnamefont {M.~S.}\ \bibnamefont
  {Wertheim}},\ }\bibfield  {title} {\bibinfo {title} {{Fluids with highly
  directional attractive forces. I. Statistical thermodynamics}},\ }\href
  {https://doi.org/10.1007/BF01017362} {\bibfield  {journal} {\bibinfo
  {journal} {J. Stat. Phys.}\ }\textbf {\bibinfo {volume} {35}},\ \bibinfo
  {pages} {19} (\bibinfo {year} {1984}{\natexlab{b}})}\BibitemShut {NoStop}%
\bibitem [{\citenamefont
  {Wertheim}(1984{\natexlab{c}})}]{Wertheim.2_JStatPhys_1984}%
  \BibitemOpen
  \bibfield  {author} {\bibinfo {author} {\bibfnamefont {M.~S.}\ \bibnamefont
  {Wertheim}},\ }\bibfield  {title} {\bibinfo {title} {{Fluids with highly
  directional attractive forces. II. Thermodynamic perturbation theory and
  integral equations}},\ }\href {https://doi.org/10.1007/BF01017363} {\bibfield
   {journal} {\bibinfo  {journal} {J. Stat. Phys.}\ }\textbf {\bibinfo {volume}
  {35}},\ \bibinfo {pages} {35} (\bibinfo {year}
  {1984}{\natexlab{c}})}\BibitemShut {NoStop}%
\bibitem [{\citenamefont
  {Wertheim}(1986{\natexlab{a}})}]{Wertheim.3_JStatPhys_1986}%
  \BibitemOpen
  \bibfield  {author} {\bibinfo {author} {\bibfnamefont {M.~S.}\ \bibnamefont
  {Wertheim}},\ }\bibfield  {title} {\bibinfo {title} {{Fluids with highly
  directional attractive forces. III. Multiple attraction sites}},\ }\href
  {https://doi.org/10.1007/BF01127721} {\bibfield  {journal} {\bibinfo
  {journal} {J. Stat. Phys.}\ }\textbf {\bibinfo {volume} {42}},\ \bibinfo
  {pages} {459} (\bibinfo {year} {1986}{\natexlab{a}})}\BibitemShut {NoStop}%
\bibitem [{\citenamefont
  {Wertheim}(1986{\natexlab{b}})}]{Wertheim.4_JStatPhys_1986}%
  \BibitemOpen
  \bibfield  {author} {\bibinfo {author} {\bibfnamefont {M.~S.}\ \bibnamefont
  {Wertheim}},\ }\bibfield  {title} {\bibinfo {title} {{Fluids with highly
  directional attractive forces. IV. Equilibrium polymerization}},\ }\href
  {https://doi.org/10.1007/BF01127722} {\bibfield  {journal} {\bibinfo
  {journal} {J. Stat. Phys.}\ }\textbf {\bibinfo {volume} {42}},\ \bibinfo
  {pages} {477} (\bibinfo {year} {1986}{\natexlab{b}})}\BibitemShut {NoStop}%
\bibitem [{\citenamefont {Wertheim}(1987)}]{Wertheim_JChemPhys_1987}%
  \BibitemOpen
  \bibfield  {author} {\bibinfo {author} {\bibfnamefont {M.~S.}\ \bibnamefont
  {Wertheim}},\ }\bibfield  {title} {\bibinfo {title} {Thermodynamic
  perturbation theory of polymerization},\ }\href
  {https://doi.org/10.1063/1.453326} {\bibfield  {journal} {\bibinfo  {journal}
  {The Journal of Chemical Physics}\ }\textbf {\bibinfo {volume} {87}},\
  \bibinfo {pages} {7323} (\bibinfo {year} {1987})}\BibitemShut {NoStop}%
\bibitem [{\citenamefont {Cuesta}\ and\ \citenamefont
  {Sanchez}(2004)}]{Cuesta-Sanchez_JStatPhys_2004}%
  \BibitemOpen
  \bibfield  {author} {\bibinfo {author} {\bibfnamefont {J.}~\bibnamefont
  {Cuesta}}\ and\ \bibinfo {author} {\bibfnamefont {A.}~\bibnamefont
  {Sanchez}},\ }\bibfield  {title} {\bibinfo {title} {{General non-existence
  theorem for phase transitions in one-dimensional systems with short range
  interactions, and physical examples of such transitions}},\ }\href
  {https://doi.org/{10.1023/B:JOSS.0000022373.63640.4e}} {\bibfield  {journal}
  {\bibinfo  {journal} {{J. Stat. Phys.}}\ }\textbf {\bibinfo {volume}
  {{115}}},\ \bibinfo {pages} {869} (\bibinfo {year} {{2004}})},\ \bibinfo
  {note} {{FisEs 2002 Meeting, Tarragona, SPAIN, 2002}}\BibitemShut {NoStop}%
\bibitem [{\citenamefont {Gurin}\ and\ \citenamefont
  {Varga}(2013)}]{Gurin-Varga_JCP_2013}%
  \BibitemOpen
  \bibfield  {author} {\bibinfo {author} {\bibfnamefont {P.}~\bibnamefont
  {Gurin}}\ and\ \bibinfo {author} {\bibfnamefont {S.}~\bibnamefont {Varga}},\
  }\bibfield  {title} {\bibinfo {title} {Pair correlation functions of two- and
  three-dimensional hard-core fluids confined into narrow pores: Exact results
  from transfer-matrix method},\ }\href {https://doi.org/10.1063/1.4852181}
  {\bibfield  {journal} {\bibinfo  {journal} {J. Chem. Phys.}\ }\textbf
  {\bibinfo {volume} {139}},\ \bibinfo {pages} {244708} (\bibinfo {year}
  {2013})}\BibitemShut {NoStop}%
\bibitem [{\citenamefont {Jackson}\ \emph {et~al.}(1988)\citenamefont
  {Jackson}, \citenamefont {Chapman},\ and\ \citenamefont
  {Gubbins}}]{Jackson-Chapman-Gubbins.1_MolPhys_1988}%
  \BibitemOpen
  \bibfield  {author} {\bibinfo {author} {\bibfnamefont {G.}~\bibnamefont
  {Jackson}}, \bibinfo {author} {\bibfnamefont {W.~G.}\ \bibnamefont
  {Chapman}},\ and\ \bibinfo {author} {\bibfnamefont {K.~E.}\ \bibnamefont
  {Gubbins}},\ }\bibfield  {title} {\bibinfo {title} {{Phase equilibria of
  associating fluids: Spherical molecules with multiple bonding sites}},\
  }\href {https://doi.org/10.1080/00268978800100821} {\bibfield  {journal}
  {\bibinfo  {journal} {Mol. Phys.}\ }\textbf {\bibinfo {volume} {65}},\
  \bibinfo {pages} {1} (\bibinfo {year} {1988})}\BibitemShut {NoStop}%
\bibitem [{\citenamefont {Paricaud}\ \emph {et~al.}(2002)\citenamefont
  {Paricaud}, \citenamefont {Galindo},\ and\ \citenamefont
  {Jackson}}]{PARICAUD200287}%
  \BibitemOpen
  \bibfield  {author} {\bibinfo {author} {\bibfnamefont {P.}~\bibnamefont
  {Paricaud}}, \bibinfo {author} {\bibfnamefont {A.}~\bibnamefont {Galindo}},\
  and\ \bibinfo {author} {\bibfnamefont {G.}~\bibnamefont {Jackson}},\
  }\bibfield  {title} {\bibinfo {title} {Recent advances in the use of the saft
  approach in describing electrolytes, interfaces, liquid crystals and
  polymers},\ }\href
  {https://doi.org/https://doi.org/10.1016/S0378-3812(01)00659-8} {\bibfield
  {journal} {\bibinfo  {journal} {Fluid Ph. Equilib.}\ }\textbf {\bibinfo
  {volume} {194-197}},\ \bibinfo {pages} {87} (\bibinfo {year} {2002})},\
  \bibinfo {note} {proceedings of the Ninth International Conference on
  Properties and Phase Equilibria for Product and Process Design}\BibitemShut
  {NoStop}%
\bibitem [{\citenamefont {Jonas}\ \emph {et~al.}(2022)\citenamefont {Jonas},
  \citenamefont {Schall},\ and\ \citenamefont {Bolhuis}}]{10.1063/5.0098882}%
  \BibitemOpen
  \bibfield  {author} {\bibinfo {author} {\bibfnamefont {H.~J.}\ \bibnamefont
  {Jonas}}, \bibinfo {author} {\bibfnamefont {P.}~\bibnamefont {Schall}},\ and\
  \bibinfo {author} {\bibfnamefont {P.~G.}\ \bibnamefont {Bolhuis}},\
  }\bibfield  {title} {\bibinfo {title} {Extended wertheim theory predicts the
  anomalous chain length distributions of divalent patchy particles under
  extreme confinement},\ }\href {https://doi.org/10.1063/5.0098882} {\bibfield
  {journal} {\bibinfo  {journal} {J. Chem. Phys.}\ }\textbf {\bibinfo {volume}
  {157}},\ \bibinfo {pages} {094903} (\bibinfo {year} {2022})}\BibitemShut
  {NoStop}%
\bibitem [{\citenamefont {Teixeira}\ and\ \citenamefont
  {Tavares}(2017)}]{Teixeira-Tavares_CurrOpinColloidInterfaceSci_2017}%
  \BibitemOpen
  \bibfield  {author} {\bibinfo {author} {\bibfnamefont {P.}~\bibnamefont
  {Teixeira}}\ and\ \bibinfo {author} {\bibfnamefont {J.}~\bibnamefont
  {Tavares}},\ }\bibfield  {title} {\bibinfo {title} {Phase behaviour of pure
  and mixed patchy colloids — theory and simulation},\ }\href
  {https://doi.org/https://doi.org/10.1016/j.cocis.2017.03.011} {\bibfield
  {journal} {\bibinfo  {journal} {Curr. Opin. Colloid Interface Sci.}\ }\textbf
  {\bibinfo {volume} {30}},\ \bibinfo {pages} {16} (\bibinfo {year}
  {2017})}\BibitemShut {NoStop}%
\bibitem [{\citenamefont {Barthes}\ \emph {et~al.}(2024)\citenamefont
  {Barthes}, \citenamefont {Bernet}, \citenamefont {Grégoire},\ and\
  \citenamefont {Miqueu}}]{10.1063/5.0180795}%
  \BibitemOpen
  \bibfield  {author} {\bibinfo {author} {\bibfnamefont {A.}~\bibnamefont
  {Barthes}}, \bibinfo {author} {\bibfnamefont {T.}~\bibnamefont {Bernet}},
  \bibinfo {author} {\bibfnamefont {D.}~\bibnamefont {Grégoire}},\ and\
  \bibinfo {author} {\bibfnamefont {C.}~\bibnamefont {Miqueu}},\ }\bibfield
  {title} {\bibinfo {title} {A molecular density functional theory for
  associating fluids in {3D} geometries},\ }\href
  {https://doi.org/10.1063/5.0180795} {\bibfield  {journal} {\bibinfo
  {journal} {J. Chem. Phys.}\ }\textbf {\bibinfo {volume} {160}},\ \bibinfo
  {pages} {054704} (\bibinfo {year} {2024})}\BibitemShut {NoStop}%
\bibitem [{\citenamefont {Marshall}\ and\ \citenamefont
  {Chapman}(2013)}]{10.1063/1.4776759}%
  \BibitemOpen
  \bibfield  {author} {\bibinfo {author} {\bibfnamefont {B.~D.}\ \bibnamefont
  {Marshall}}\ and\ \bibinfo {author} {\bibfnamefont {W.~G.}\ \bibnamefont
  {Chapman}},\ }\bibfield  {title} {\bibinfo {title} {A density functional
  theory for patchy colloids based on wertheim's association theory: Beyond the
  single bonding condition},\ }\href {https://doi.org/10.1063/1.4776759}
  {\bibfield  {journal} {\bibinfo  {journal} {J. Chem. Phys.}\ }\textbf
  {\bibinfo {volume} {138}},\ \bibinfo {pages} {044901} (\bibinfo {year}
  {2013})}\BibitemShut {NoStop}%
\bibitem [{\citenamefont {Algaba}\ \emph {et~al.}(2022)\citenamefont {Algaba},
  \citenamefont {Mendiboure}, \citenamefont {Gómez-Álvarez},\ and\
  \citenamefont {Blas}}]{D2RA02162E}%
  \BibitemOpen
  \bibfield  {author} {\bibinfo {author} {\bibfnamefont {J.}~\bibnamefont
  {Algaba}}, \bibinfo {author} {\bibfnamefont {B.}~\bibnamefont {Mendiboure}},
  \bibinfo {author} {\bibfnamefont {P.}~\bibnamefont {Gómez-Álvarez}},\ and\
  \bibinfo {author} {\bibfnamefont {F.~J.}\ \bibnamefont {Blas}},\ }\bibfield
  {title} {\bibinfo {title} {Density functional theory for the prediction of
  interfacial properties of molecular fluids within the saft-$\gamma$
  coarse-grained approach},\ }\href {https://doi.org/10.1039/D2RA02162E}
  {\bibfield  {journal} {\bibinfo  {journal} {RSC Adv.}\ }\textbf {\bibinfo
  {volume} {12}},\ \bibinfo {pages} {18821} (\bibinfo {year}
  {2022})}\BibitemShut {NoStop}%
\bibitem [{\citenamefont {McGrother}\ \emph {et~al.}(1997)\citenamefont
  {McGrother}, \citenamefont {Sear},\ and\ \citenamefont
  {Jackson}}]{10.1063/1.473693}%
  \BibitemOpen
  \bibfield  {author} {\bibinfo {author} {\bibfnamefont {S.~C.}\ \bibnamefont
  {McGrother}}, \bibinfo {author} {\bibfnamefont {R.~P.}\ \bibnamefont
  {Sear}},\ and\ \bibinfo {author} {\bibfnamefont {G.}~\bibnamefont
  {Jackson}},\ }\bibfield  {title} {\bibinfo {title} {The liquid crystalline
  phase behavior of dimerizing hard spherocylinders},\ }\href
  {https://doi.org/10.1063/1.473693} {\bibfield  {journal} {\bibinfo  {journal}
  {J. Chem. Phys.}\ }\textbf {\bibinfo {volume} {106}},\ \bibinfo {pages}
  {7315} (\bibinfo {year} {1997})}\BibitemShut {NoStop}%
\bibitem [{\citenamefont {Lebowitz}\ and\ \citenamefont
  {Zomick}(1971)}]{Lebowitz-Zomick_JChemPhys_1971}%
  \BibitemOpen
  \bibfield  {author} {\bibinfo {author} {\bibfnamefont {J.~L.}\ \bibnamefont
  {Lebowitz}}\ and\ \bibinfo {author} {\bibfnamefont {D.}~\bibnamefont
  {Zomick}},\ }\bibfield  {title} {\bibinfo {title} {Mixtures of hard spheres
  with nonadditive diameters: Some exact results and solution of py equation},\
  }\href {https://doi.org/10.1063/1.1675348} {\bibfield  {journal} {\bibinfo
  {journal} {J. Chem. Phys.}\ }\textbf {\bibinfo {volume} {54}},\ \bibinfo
  {pages} {3335} (\bibinfo {year} {1971})}\BibitemShut {NoStop}%
\bibitem [{\citenamefont {Gil-Villegas}\ \emph {et~al.}(1997)\citenamefont
  {Gil-Villegas}, \citenamefont {Galindo}, \citenamefont {Whitehead},
  \citenamefont {Mills}, \citenamefont {Jackson},\ and\ \citenamefont
  {Burgess}}]{10.1063/1.473101}%
  \BibitemOpen
  \bibfield  {author} {\bibinfo {author} {\bibfnamefont {A.}~\bibnamefont
  {Gil-Villegas}}, \bibinfo {author} {\bibfnamefont {A.}~\bibnamefont
  {Galindo}}, \bibinfo {author} {\bibfnamefont {P.~J.}\ \bibnamefont
  {Whitehead}}, \bibinfo {author} {\bibfnamefont {S.~J.}\ \bibnamefont
  {Mills}}, \bibinfo {author} {\bibfnamefont {G.}~\bibnamefont {Jackson}},\
  and\ \bibinfo {author} {\bibfnamefont {A.~N.}\ \bibnamefont {Burgess}},\
  }\bibfield  {title} {\bibinfo {title} {Statistical associating fluid theory
  for chain molecules with attractive potentials of variable range},\ }\href
  {https://doi.org/10.1063/1.473101} {\bibfield  {journal} {\bibinfo  {journal}
  {J. Chem. Phys.}\ }\textbf {\bibinfo {volume} {106}},\ \bibinfo {pages}
  {4168} (\bibinfo {year} {1997})}\BibitemShut {NoStop}%
\end{thebibliography}%

\end{document}